\documentclass[pre,amsmath]{revtex4-2}
\usepackage[pdftex]{graphicx} 
\usepackage{color}
\usepackage{bm}
\usepackage{epsfig}
\usepackage{latexsym}
\usepackage{nameref,hyperref}
\begin{document}
\newcommand{\be}{\begin{equation}}
\newcommand{\ee}{\end{equation}}
\newcommand{\bea}{\begin{eqnarray}}
\newcommand{\eea}{\end{eqnarray}}
\newcommand{\RN}[1]{\textup{\uppercase\expandafter{\romannumeral#1}}}
\title{Effect of receptor cooperativity on methylation dynamics in bacterial chemotaxis with weak and strong gradient}
\author{Shobhan Dev Mandal and Sakuntala Chatterjee}
\affiliation{Department of Theoretical Sciences, S. N. Bose National Centre for Basic Sciences, Block JD, Sector 3, Salt Lake, Kolkata 700106, India.}
\begin{abstract}
We study methylation dynamics of the chemoreceptors as an {\sl E.coli} cell moves around in a spatially varying chemo-attractant environment. We consider attractant concentration with strong and weak spatial gradient. During the uphill and downhill motion of the cell along the gradient, we measure the temporal variation of average methylation level of the receptor clusters. Our numerical simulations show that the methylation dynamics depends sensitively on the size of the receptor clusters and also on the strength of the gradient. At short times after the beginning of a run, the methylation dynamics is mainly controlled by short runs which are generally associated with high receptor activity. This results in demethylation at short times. But for intermediate or large times, long runs play an important role and depending on receptor cooperativity or gradient strength, the qualitative variation of methylation can be completely different in this time regime. For weak gradient, both for uphill and downhill runs, after the initial demethylation, we find methylation level increases steadily with time for all cluster sizes. Similar qualitative behavior is observed for strong gradient during uphill runs as well. However, the methylation dynamics for downhill runs in strong gradient show highly non-trivial dependence on the receptor cluster size. We explain this behavior as a result of interplay between the sensing and adaptation modules of the signaling network.
\end{abstract}
\maketitle
\section{Introduction}

An {\sl E.coli} cell uses run-and-tumble motion to climb up the concentration gradient of a nutrient or a chemical attractant \cite{berg1972chemotaxis, adler1973method, adler1973chemotaxis}. This directed migration, guided by chemical environment surrounding the cell, is known as chemotaxis \cite{berg2008coli}. The signaling network inside an {\sl E.coli} cell consists of two principal modules: sensing and adaptation \cite{tu2013quantitative}. These two modules are coupled to each other via the activity of the chemoreceptors. When the transmembrane chemoreceptors bind to  the attractant molecules, their activity decreases. This input signal is processed by the sensing module of the network, and the run-and-tumble motion of the cell is in turn modulated such that the cell shows a net drift towards the region with higher  attractant concentration \cite{chatterjee2011chemotaxis, dev2018optimal, vladimirov2010predicted}. The adaptation part of the network implements a negative feedback mechanism to ensure that the receptor activity does not get too high or too low. This is done by controlling the methylation level of the chemoreceptor such that the activity level is restored to its  adapted value. An important aspect of the chemotactic signaling network is the cooperativity or clustering tendency of the chemoreceptors, which allows them to form clusters or `signaling teams' \cite{frank2016networked}. All receptors in a team change their activity in unison which results in significant amplification of the input signal. This makes it possible for an {\sl E.coli} to sense even weak concentration gradient of the  attractant \cite{duke1999heightened, bray1998receptor}.

Recent experimental and theoretical studies \cite{keegstra2017phenotypic, colin2017multiple, shobhan} have shown that receptor clustering is also an important source of intracellular fluctuation. In other words, enhanced sensitivity comes at the cost of increased biochemical noise. This results in an optimum size of the receptor cluster, or equivalently, an optimum strength of the cooperative interaction between the receptors, for which the cell shows the most efficient chemotaxis. The origin of this optimality is distinctly different from the optimal signaling team size observed for a noisy input signal, when due to amplification of that noise the chemotactic performance gets adversely affected for very large signaling teams \cite{aquino2011optimal}. We had shown in an earlier work \cite{shobhan} that even in absence of any noise in the ligand environment, there exists an optimal size of the signaling team when the chemotactic performance is at its best. We had calculated different quantities characterizing the chemotactic performance and shown how each of them reaches a peak at a specific size of the signaling team \cite{shobhan}.

In the present work, we investigate how the cooperativity of the chemoreceptors affects their methylation levels. More precisely, we study the temporal variation of the receptor methylation levels as the cell navigates through spatially varying attractant environment. Our extensive numerical simulations on a detailed theoretical model \cite{pontius2013adaptation, shobhan} show that the nature of variation depends sensitively on the level of cooperativity present among the receptors. We explicitly consider two different types (directions) of runs: uphill and downhill. As the names suggest, during an uphill run the local attractant concentration increases along the cell trajectory and during a downhill run it decreases. We monitor the average change in methylation level of a receptor cluster as a function of time during these runs.

Although there have been a lot of studies on {\sl E.coli} chemotaxis  \cite{wu1996receptor, vladimirov2008dependence, matthaus2009coli, vladimirov2010predicted, jiang2010quantitative, matthaus2011origin}, very little is known about how the methylation level of the chemoreceptors vary with time as the cell samples different attractant concentrations during its run-and-tumble motion. Due to the complex nature of the signaling network, it is not straightforward to predict this temporal variation. The chemoreceptor activity which controls the tumbling bias, depends on local ligand concentration through the sensing module of the network: an increase (decrease) in ligand concentration tends to reduce (raise) the activity. Through the adaptation module of the network, activity also depends on the methylation level: active (inactive) receptors get demethylated (methylated) which in turn lowers (raises) the activity. Because of this coupled dynamics, the temporal variation of methylation level can be quite complicated. Our study unravels this complex dynamics and provides useful insights into the interdependence of different dynamical variables in the signaling network.

Our numerical simulations show how methylation dynamics is controlled by the direction and strength of the concentration gradient of the attractant, as well as the receptor cooperativity. Defining time $t=0$ at the start of an uphill or a downhill run, we monitor the change in average methylation level as a function of time $t$ during those runs. We mainly use two different quantities to measure the above change. At time $t$ we calculate the average methylation level for all those runs which persist ({\sl i.e.} do not tumble) till time $t$ and subtract from it the initial average methylation level of those same set of runs. Or alternatively, we can subtract the average initial methylation levels of all runs starting at $t=0$, irrespective of their durations. We find the choice of initial ensemble is crucial and can lead to qualitatively different nature of temporal variations. This difference is particularly pronounced for small sizes of the receptor clusters. Our data also show that the interplay between the sensing and adaptation modules of the signaling network manifests itself in different ways depending on the strength of the attractant concentration gradient and on the size of the receptor clusters. This interplay strongly affects the temporal variation of the methylation levels. We explain our numerical observations from a detailed analysis of the coupled time evolution of receptor activity, methylation and ligand concentration within the signaling network. To the best of our knowledge, such a systematic investigation of methylation dynamics along the cell trajectory has never been done before. Our study reveals a rich methylation dynamics and its sensitive dependence on receptor cooperativity and gradient strength.

We also suggest possible experiments to verify our conclusions. Note that our study is based on a swimming cell, whereas most experiments are performed on tethered cell. Therefore, we propose an experimental protocol involving a tethered cell and a time-varying  attractant level \cite{shimizu2010modular} that mimics the situations experienced by a swimming cell. In the next section of the paper we present our model. In Sec \ref{sec:result} we present our results on methylation dynamics for weak and strong gradient cases in two different subsections. In Sec \ref{sec:con} we summarize our results, discuss their significance and suggest possible experimental verification. In appendices \ref{app:model}-\ref{app:expt} we present additional details of model, parameters, simulations and protocol of suggested experiments. We also present additional supporting data in appendices \ref{app:act}-\ref{app:2d}.

\section{Model description}
\label{sec:model}

There are few thousand chemoreceptors in an {\sl E.coli} cell which exist in the form of dimers \cite{briegel2012bacterial, liu2012molecular}. Each receptor dimer can be in two different states: active or inactive. According to Monod-Wyman-Changeux (MWC) model \cite{mello2005allosteric, keymer2006chemosensing, monod1965nature}, the free energy difference between these two states (in the units of $K_B T$) can be written as 
\begin{equation}
\epsilon [ c(x), m] = \log \frac{1+c(x)/K_{min}}{1+c(x)/K_{max}} + \epsilon_0 - \epsilon_1 m
\label{eq:mwc}
\end{equation}
where $x$ denotes the location of the cell and $c(x)$ is the concentration of the  attractant at that location, $m$ stands for the methylation level of the dimer which can take any integer value between $0$ and $8$  \cite{pontius2013adaptation, dufour2014limits, frankel2014adaptability, long2017feedback} and $\epsilon_0$ and $\epsilon_1$ are two constants whose values are listed in Table \ref{table}.  The parameters $K_{min}$ and $K_{max}$ set the range of sensitivity, {\sl i.e.} the cell can sense any concentration as long as $K_{min} < c(x) < K_{max}$ is satisfied. A receptor cluster of size $n$ contains $3n$ dimers, where the factor $3$ accounts for the formation of trimer of dimers. The total free energy of the cluster is the sum of individual free energies of all $3n$ dimers
\be
F = 3n  \log \frac{1+c(x)/K_{min}}{1+c(x)/K_{max}}  +3n \epsilon_0  - \sum_{i=1} ^{3n} \epsilon_1 m(i) 
\label{eq:fcl}
\ee
where $m(i)$ is the methylation level of the $i$-th dimer. Due to cooperativity, all dimers in one cluster change their activity state simultaneously and this transition probability depends on the free energy of the cluster (see Appendix \ref{app:model} for more details). The activity of a cluster is defined as the probability to find the cluster in the active state and its long time average is given by $[1+\exp(F)]^{-1}$. The total activity of the cell is obtained by averaging over activities of all clusters. When clusters are larger in size, they are fewer in number and the total activity is then a result of averaging over a smaller number of clusters. This is why activity fluctuations increase for large $n$ \cite{shobhan, colin2017multiple}.

The total activity determines the tumbling bias of the cell. In the run mode the cell swims smoothly, and it can tumble with a certain probability which takes a large value if the activity is high. In the tumble mode the cell undergoes random rotation about its body axis when it reorients itself and can switch back to the run mode again \cite{dev2015search, dev2019rnt}. The tumble to run switching rate is high if the activity is low. During the motion of the cell its position $x$ changes and hence $c(x)$  changes. It follows from Eq. \ref{eq:mwc} that if $c(x)$ increases, activity decreases and runs tend to get longer. Similarly, for decreasing $c(x)$ runs tend to get shorter. This gives rise to a net migration up the gradient of $c(x)$. In this work, we have considered a linear concentration profile: $c(x) = c_0 (1+x/x_0)$ where $c_0$ is the background concentration and $x_0$ measures the strength of the gradient.  We perform agent-based simulations where we consider the movement of the cell in both one and two dimensions with reflecting boundary walls. In the main paper we present results for the one dimensional motion of the cell, and in Appendix \ref{app:2d} we present the two dimensional case. Our main conclusions remain same in both cases.

The other dynamical variable in Eq. \ref{eq:mwc} is $m$, which is controlled by methylating enzyme CheR and demethylating enzyme CheB-P. A dimer can bind to one enzyme molecule at a time. An active dimer gets demethylated by CheB-P which lowers its $m$ value by $1$, provided $m >0$. Similarly, an inactive dimer gets methylated by CheR which raises its $m$ by $1$, provided $m <8$. From Eq. \ref{eq:fcl} it follows that methylation increases activity and demethylation decreases it. This constitutes the negative feedback which is responsible for adaptation in the network. In Appendix \ref{app:model} we provide details of the binding-unbinding kinetics of the enzyme molecules. One important point that needs to be considered here, is the very low concentration of enzyme molecules, as compared to very large number of receptor dimers \cite{li2004cellular}. Because of this the methylation level of a dimer changes very slowly and in fact methylation-demethylation process is the slowest step in the whole reaction network. However, this does not adversely affect the adaptation capability of the cell and {\sl E.coli} cell is known to show near-perfect adaptation \cite{berg1975transient, goy1977sensory}. To explain this a number of mechanisms have been proposed experimentally and theoretically, like assistance neighborhood and brachiation \cite{levin2002binding, endres2006precise, hansen2008chemotaxis, kim2002dynamic, li2005adaptational}. In an assistance neighborhood model, one enzyme molecule can tether to one receptor dimer and can modify the methylation levels of the dimers in its neighborhood. In brachiation model, the enzyme molecule, once bound to a dimer can perform random walk on the receptor array and move from one dimer to others and modify their methylation levels. We include a flavor of these mechanisms in our model, as explained in more details in Appendix \ref{app:model}. We list the values of all parameters we used in the Table \ref{table}.

\section{Temporal variation of methylation levels}
\label{sec:result}

In this section, we present our results on temporal variation of methylation level of a receptor cluster during uphill and downhill runs of the cell.  We consider a large enough number of uphill and downhill runs of variable durations and average over them to measure the methylation variation. Let $N^+(t)$ and $N^-(t)$ be the number of uphill and downhill runs, respectively whose duration is larger than $t$ where $t=0$ is set at the start of the run.  Our data in Fig. \ref{fig:runud} show that $N^\pm (t)$ decrease exponentially with time. Let $m_i^+(t)$ be the methylation level of a particular receptor cluster measured at time $t$ during the $i$-th uphill run. In a similar way, $m_i^-(t)$ can also be defined.  We define the following quantities
\be
\Delta m ^\pm (t) = \sum_{i=1}^{N^\pm (t)} \frac{m_i ^\pm (t)-m_i ^\pm (0)}{N^\pm (t)}
\label{eq:pers}
\ee
and 
\begin{equation}
\delta m ^\pm (t) = \frac{\sum_{i=1}^{N^\pm (t)} m_i^\pm (t)}{N^\pm (t)} - \frac{\sum_{i=1}^{N^\pm (0)} m_i^\pm(0)}{N^\pm (0)}
\label{eq:all}
\end{equation}
where all upper (lower) signs in the superscripts correspond to uphill (downhill) runs. $\Delta m^+ (t)$ ($\Delta m ^-(t)$) considers those uphill (downhill) runs which persist at least till time $t$ and measures the average change in methylation in those runs. On the other hand, $\delta m^+ (t)$ ($\delta m ^ -(t)$) considers average methylation level of all uphill (downhill) runs persistent till time $t$ and subtracts from it the average methylation level of all uphill (downhill) runs which started at time $t=0$, irrespective of their durations.  More simply, $\dfrac{\sum_{i=1}^{N^\pm (t)} m_i^\pm (t)}{N^\pm (t)} $ is a quantity that tracks average methylation level of a running cell at time $t$ and just by subtracting the initial value of this quantity gives us $ \delta m ^\pm (t)$. As clear from Eqs. \ref{eq:pers} and \ref{eq:all} the main difference between $\Delta m^\pm (t)$ and $\delta m ^\pm (t)$ is in the choice of the initial averaging at $t=0$. For $\Delta m^\pm (t)$ we subtract the average initial methylation of only those runs which persist till $t$, while for $\delta m^\pm (t)$ we subtract average initial methylation of all runs. As we show below, this is an important difference and can result in completely different nature of time-dependence of these two quantities. We are interested in the effect of receptor cooperativity on the temporal variation of these quantities. We find the effect is very different for weak gradient case and strong gradient case. We separately present these two cases below. { For comparison, we also show results for the homogeneous attractant environment for which $\Delta m^+(t)$ and $\Delta m^-(t) $ become identical. Similarly, $\delta m^+(t)$ and $\delta m^-(t)$ also become same in this limit. One might expect that the methylation variation for the homogeneous environment lies in between the uphill and downhill variation. But we show below that it is not always true. }

\subsection{Weak gradient}
\label{sec:weak}
\begin{figure}
\includegraphics[scale=1.2]{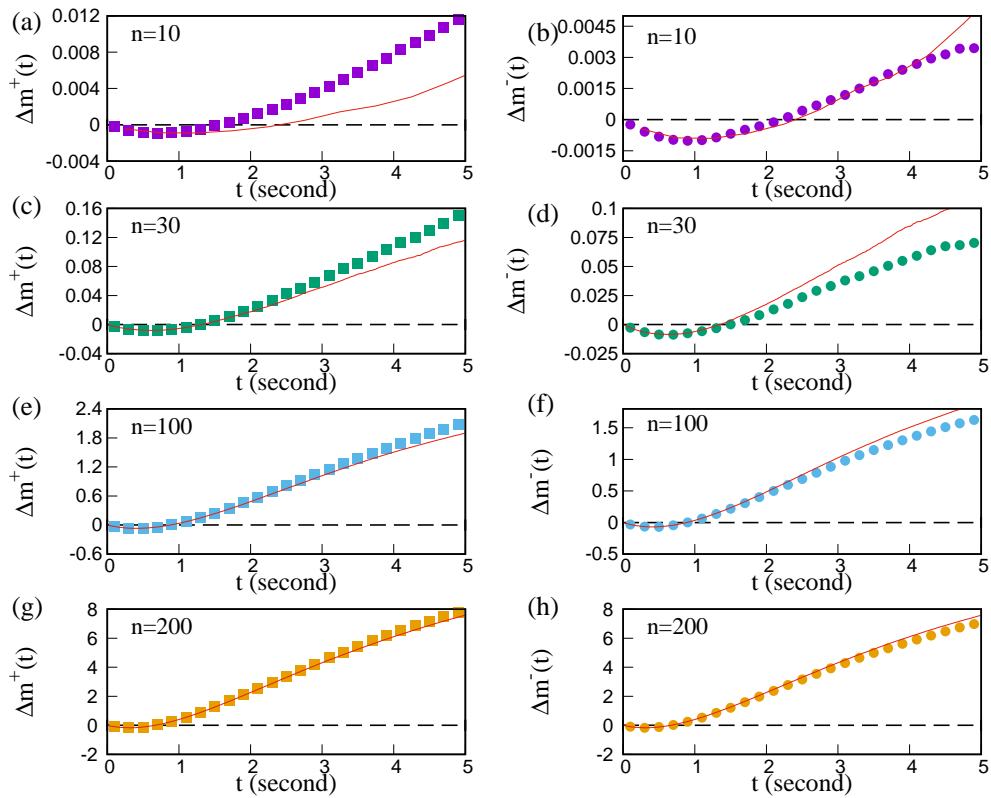}
\caption{Temporal variation of $\Delta m^\pm (t)$ for different $n$: left panel shows plots for $\Delta m^+(t)$ for the uphill runs and the right panel shows $\Delta m^-(t)$ for the downhill runs. Initial demethylation is due to short high activity runs and later methylation is due to long runs with low activity. These data are for a one dimensional motion of the cell in a box of size $L$ across which a linear concentration profile $c(x)$ of the  attractant is set up with weak gradient. The red solid lines represent the data for a homogeneous attractant environment. All simulation parameters are listed in Table \ref{table} in Appendix \ref{app:model}. These data are averaged over at least $8 \times 10^5$ histories.}
\label{fig:Dmw}
\end{figure}

For weak gradient we consider $x_0 = 20$ $mm$. In this case we expect the long time distribution of the cell position inside the box to be well-approximated by a linearly varying function. We present our data for $\Delta m^\pm (t)$ in Fig. \ref{fig:Dmw} for different values of the receptor cluster size $n$. 

\begin{figure}
\includegraphics[scale=0.7,angle=270 ]{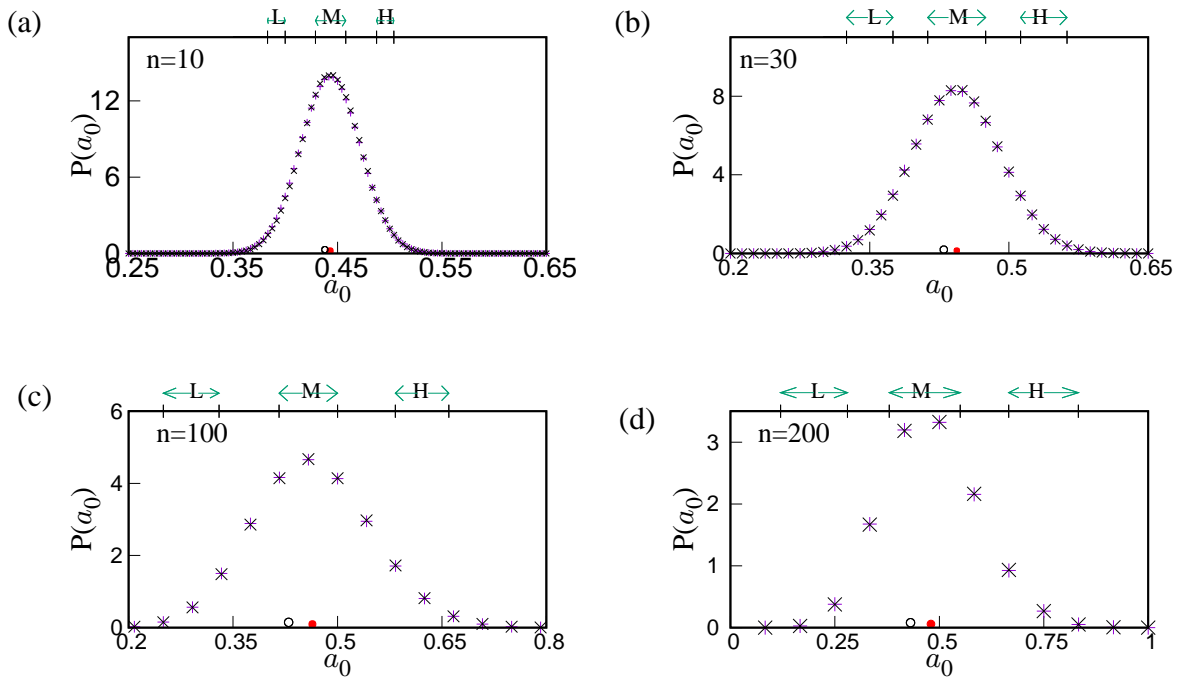}
\caption{ Distribution of activity $a_0$ at the start of a run.  The plus symbols show data for weak gradient. The cross symbols are for the zero gradient case, which almost overlap with the weak gradient data. The low (L), medium (M) and high (H) ranges of values of $a_0$ have been shown. These ranges are defined with reference to the mean $a_0$ value shown by the red point. The empty circle on the $x$-axis shows the adapted activity value which also belongs to the medium range. Each data point has been averaged over at least $10^6$ histories. All simulation parameters are listed in Table \ref{table} in Appendix \ref{app:model}. }
\label{fig:a0distw}
\end{figure}

To analyze these data, it is useful to consider separately the runs starting with different ranges of activity values. To this end, we show in Fig. \ref{fig:a0distw} the distribution of activity $a_0$ values at the beginning of a run. The distribution is unimodal and gets wider with increasing $n$, which is consistent with earlier results known for steady state activity distribution \cite{shobhan, keegstra2017phenotypic, colin2017multiple}. The distribution is also roughly symmetric about the peak. The mean value of $a_0$ (shown by a red filled circle in Fig. \ref{fig:a0distw}), as well as the adapted level of activity (shown by an empty circle), lie close to the peak position. Therefore, the activity values near the peak can be considered to be in the medium range, the ones near the left (right) tail belong to low (high) range. In Fig. \ref{fig:rundist} of Appendix \ref{app:lowhigh} we show the distribution of run durations which start with low, medium or high activity values. As expected, the long runs are least probable for high $a_0$ and most probable for low $a_0$. The integrated quantity, {\sl i.e.} the number of runs persisting at least till time $t$, starting with $a_0$ values belonging to these three different ranges has been plotted in Fig. \ref{fig:npm}. As we show below, these variations play a crucial role in $\Delta m^\pm (t)$ behavior.

\begin{figure}
\includegraphics[scale=0.7,angle=270]{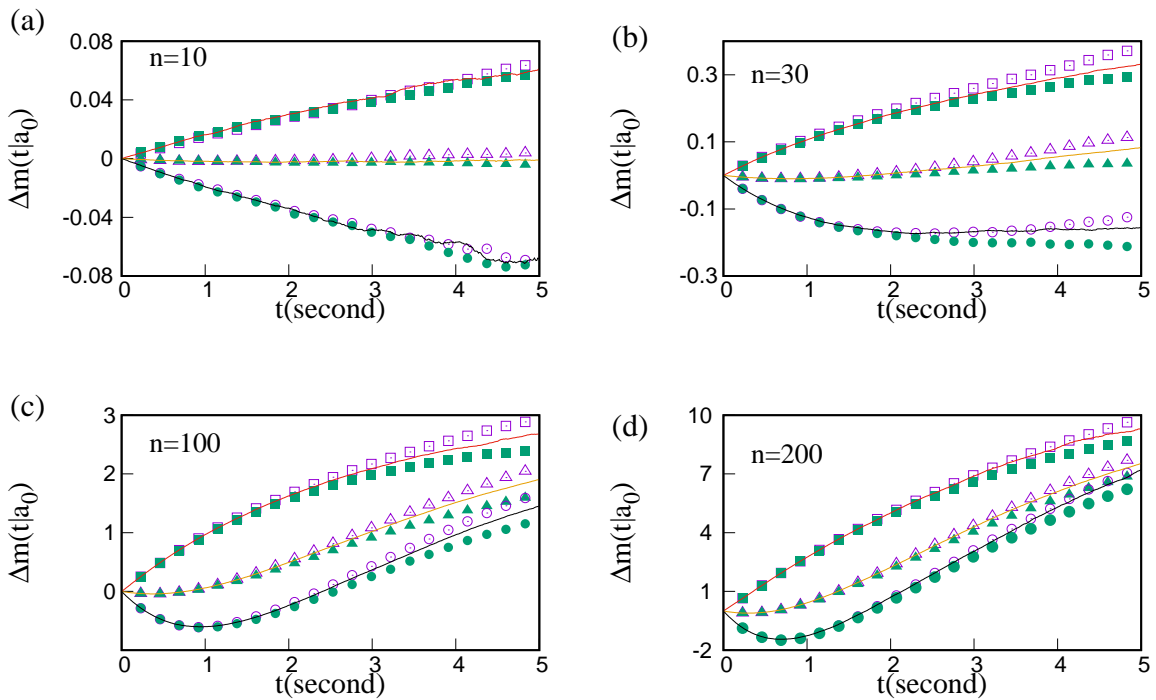}
\caption{ $\Delta m^\pm (t|a_0))$ for runs starting with three different activity ranges. The discrete points present data for the weak  gradient case and the solid lines are for the flat concentration profile of the attractant. The square, triangular and circular symbols correspond to low, medium and high $a_0$ values while the empty (filled) symbols are for uphill (downhill) runs. Among the solid lines, the top (red), middle (yellow) and the bottom (black) ones correspond to low, medium and high $a_0$ runs for the flat profile. Each data point has been averaged at least $10^5$ histories. All simulation parameters are listed in Table \ref{table} in Appendix \ref{app:model}}
\label{fig:Dmwa0}
\end{figure}

In Fig. \ref{fig:Dmwa0} we plot $\Delta m^\pm (t|a_0))$, defined as the methylation variation during a run starting with low, medium or high $a_0$. We expect that for small $a_0$ we should have methylation, for large $a_0$ we should have demethylation and for medium $a_0$ change in methylation should have small magnitude. This is exactly what we find for small $n$, as shown in Fig. \ref{fig:Dmwa0}(a). For large $n$ the behavior is same as above for small $t$ but for large $t$ our data in panels (c) and (d) of Fig. \ref{fig:Dmwa0} show positive values of  $\Delta m^\pm (t|a_0))$ which indicates methylation. This is consistent with our data in Fig. \ref{fig:Dmw}(e)-(h). It may appear counter-intuitive why for large $a_0$ even downhill runs show methylation at large times. Actually, the initial strong demethylation lowers the activity significantly since for large $n$ flipping of activity states of even few signaling teams can cause the activity value to fall below its adapted level. This triggers methylation for large $t$. In Fig. \ref{fig:Dmwa0} we also show the data for flat attractant profile. As expected, it lies between the uphill and downhill curves. 

Aided by the insights obtained from looking at the methylation variation for different activity range, we can now explain the data for $\Delta m^+(t)$ in Fig. \ref{fig:Dmw}. For $n=10$ Fig. \ref{fig:Dmw}(a) shows that  $\Delta m^+(t)$ starts from $0$, decreases slightly to become negative for small $t$ and then increases steadily with $t$. This means for small $t$ the uphill runs show demethylation and then they switch over to methylation. Note that our data in Fig. \ref{fig:Dmwa0}(a) show that the demethylation trend for high $a_0$ runs are stronger than the methylation trend for low $a_0$ runs. This is also consistent with our observation that mean value of $a_0$ is higher than the adapted activity (see Fig. \ref{fig:a0distw})(a). In other words, the short time behavior of $\Delta m^+(t)$ is dominated by high $a_0$ runs which undergo demethylation.  However, these runs are short and  as time goes on, these short runs end and drop out of $N^+(t)$ population and only the longer runs remain, as seen in Fig. \ref{fig:npm}(a). These runs are associated with low activity and hence receptor methylation. Moreover, due to increasing $c(x)$ along the cell trajectory in this case, activity is lowered further. Thus methylation takes over and $\Delta m^+(t)$ shows positive growth. In Fig. \ref{fig:Dmw}b we show the data for $\Delta m^- (t)$ for the same cluster size. The qualitative behavior remains same here also. However, the late time growth due to methylation is much weaker in this case. This is expected since even for those runs with low activity which persist till late times, $c(x)$ keeps decreasing with $t$ which tends to raise the activity. Because of this opposing effect coming from the ligand free energy, activity remains higher than the uphill runs, and average methylation shows a slower growth. To filter out the effect of the ligand concentration gradient, in Fig. \ref{fig:Dmw} we also plot the methylation variation in absence of a gradient (red lines). As expected, we find the curve in this case lies in between $\Delta m^+(t)$ and $\Delta m^-(t)$.

For larger values of $n$, qualitative behavior remains the same, although the quantitative variation of $\Delta m^\pm (t)$ happens over a larger range now. This is expected since the number of receptors per cluster increases, the total change in methylation level of a cluster also increases. The negative values observed at small $t$ also show similar trend, the minimum in $\Delta m^\pm (t)$ at small $t$ becomes deeper as $n$ increases. We also notice that for $n=10,30$ the range of variation of $\Delta m^+(t)$ and $\Delta m^-(t)$ are significantly different, but for $n=100,200$ the ranges are not as different for the uphill and downhill runs. Methylation is still slower for the downhill curve, but the values are much closer to the uphill curve. This is because for large $n$ the activity fluctuations increase and then adaptation plays a bigger role in the signaling network and ligand free energy becomes less important \cite{colin2017multiple,shobhan}. The difference in the cell behavior during an uphill and downhill run therefore decreases for large $n$. In Appendix \ref{app:act} we show this explicitly by plotting the difference between average activity during an uphill and downhill run, and as expected, this difference decreases with $n$ for large $n$. This explains why the temporal variation of methylation follows a similar course for large $n$, irrespective of whether the cell is running uphill or downhill.

\begin{figure}
\includegraphics[scale=1.2]{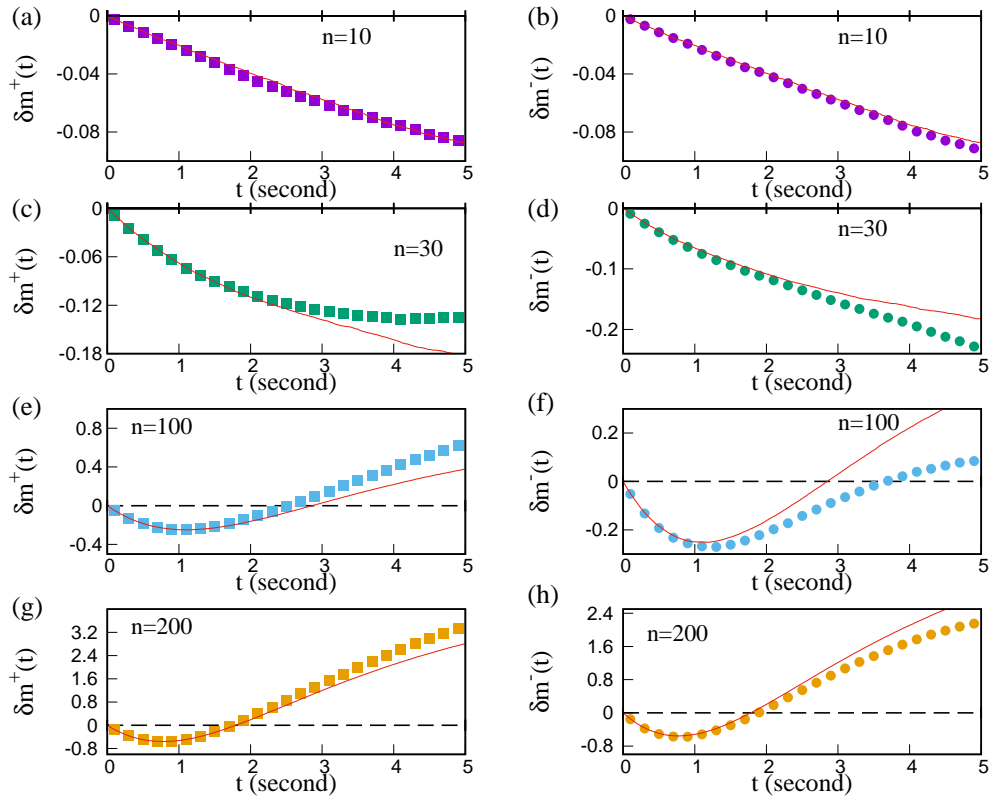}
\caption{Temporal variation of $\delta m^\pm (t)$ for different $n$: left (right) panel corresponds to uphill (downhill) runs. Increasing methylation level for long runs tends to increase average methylation, while dropping out of high methylation trajectories from $N^\pm (t)$ tends to decrease average methylation. In this competition the former wins for small $n$ and the later wins for large $n$. These data are averaged over at least $5 \times 10^5$ histories. Other simulation details are as in Fig. \ref{fig:Dmw}.}
\label{fig:dmw}
\end{figure}

Surprisingly, the temporal variation of $\delta m^\pm (t)$ shows a completely different behavior. Our data for $n=10$ in Fig. \ref{fig:dmw}  show that  $\delta m^\pm (t)$ starts from $0$ and then it decreases with time for both uphill and downhill runs. This indicates uphill and downhill runs show demethylation on an average, which is opposite to what we had seen in Fig. \ref{fig:Dmw} for the same cluster size. The reason behind this apparently contradicting observation is explained below. Since more and more runs terminate with time, $N^\pm (t)$ decreases with $t$. Since the tumbling bias increases with activity, it follows that majority of those terminated runs correspond to high activity. Our data in Fig. \ref{fig:npm}(a)  clearly show this trend, where high activity runs show sharpest decline at small $t$.  A high activity state is in turn associated with high methylation level, as follows from Eq. \ref{eq:fcl}. Therefore, those runs which drop out of $N^\pm (t)$ population, have high methylation level. Although positive values of $\Delta m ^\pm (t)$ for moderate or large $t$ ensure that runs which persist till such times undergo methylation, due to dropping out of high methylation states from the population, the average methylation level still decreases with time, making $\delta m^\pm (t)$ negative. Our data also show that $\delta m^-(t)$ takes a larger negative value at late times, compared to $\delta m ^+(t)$. This is consistent with the fact that the positive growth of $\Delta m^-(t)$ is slower than that of $\Delta m^+(t)$ for large $t$ (see Figs. \ref{fig:Dmw}(a) and \ref{fig:Dmw}(b)).  

For $n=30$ the trend remains similar, except $\delta m^+(t)$ after an initial decrease tends to saturate at larger $t$. This is because $\Delta m^+(t)$ for this $n$ shows a strong growth for large $t$ (see Fig. \ref{fig:Dmw}(c)) and although high methylation states continue to drop out of the population of $N^+(t)$, due to large rise in methylation level during the individual long runs, the decrease of average methylation level of all uphill runs gets arrested. This effect is even more prominent for $n=100, 200$ where due to even stronger growth of $\Delta m ^\pm (t)$, we find a trend reversal: $\delta m^\pm (t)$ after an initial decrease, show upward swing and start increasing with time. Here the decreasing tendency of population averaged methylation level due to tumbling is overcompensated by large growth in methylation in individual persistent runs. As expected, this effect is stronger for the uphill runs, and somewhat weaker for the downhill runs.

\subsection{Strong gradient}

\begin{figure}
\includegraphics[scale=1.2]{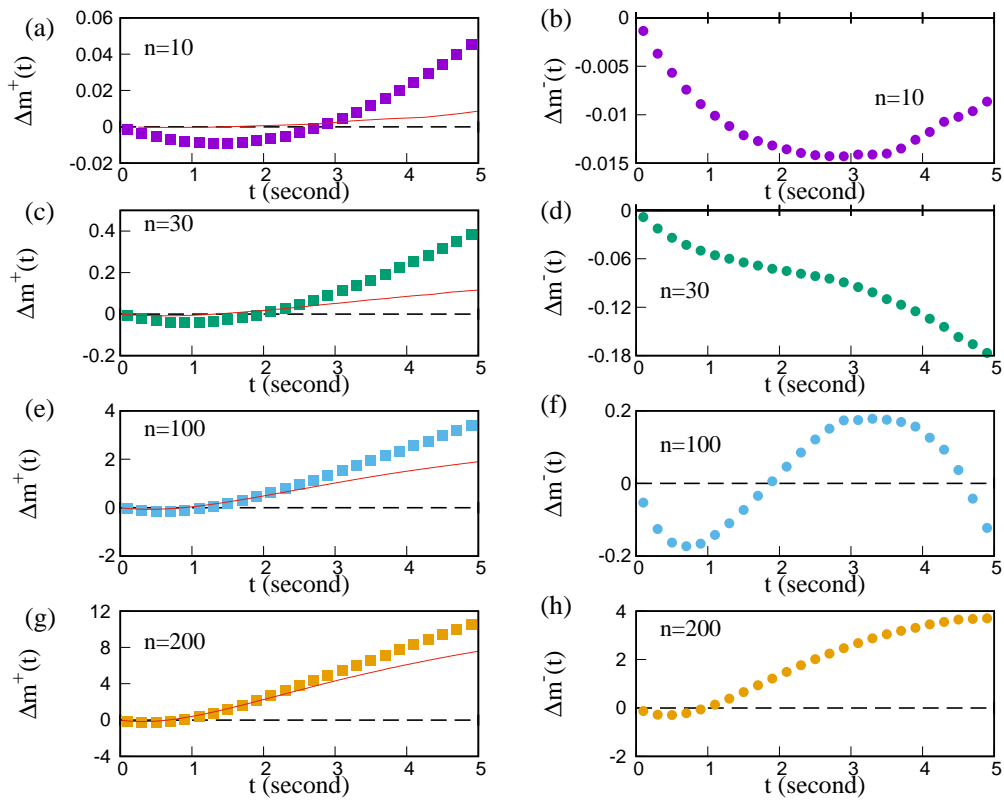}
\caption{Temporal variation of $\Delta m^\pm (t)$ for different $n$ and strong gradient case: left (right) panel corresponds to uphill (downhill) runs. While $\Delta m^+(t)$ shows qualitatively similar time-dependence as in the weak gradient case, the behavior of $\Delta m^-(t)$ is completely different. Unlike the weak gradient case, $\Delta m^-(t)$ shows significantly varying dynamics depending on the value of $n$. Such behavior is a result of interplay between the sensing module and adaptation module of the signaling network. The red lines show the data for the zero gradient case. We have here presented the data for zero gradient only in the left panel and skipped inclusion of the same data in the right panel for convenience in the choice of scales. Here, one dimensional motion of the cell is considered in presence of a strong gradient of $c(x)$. Other simulation parameters can be read off from Table \ref{table} in Appendix \ref{app:model}. These data have been averaged over at least $10^5$ histories. } 
\label{fig:Dms}
\end{figure}

To investigate the case of strong gradient we use $x_0 = 2 mm$, which is ten times stronger than what we had considered in the previous subsection. The steady state distribution of the cell position in this case has an exponential form. We present our data for $\Delta m^\pm (t)$ in Fig. \ref{fig:Dms}. We find qualitatively different dynamics compared to the weak gradient case. 

\begin{figure}
\includegraphics[scale=0.7,angle=270 ]{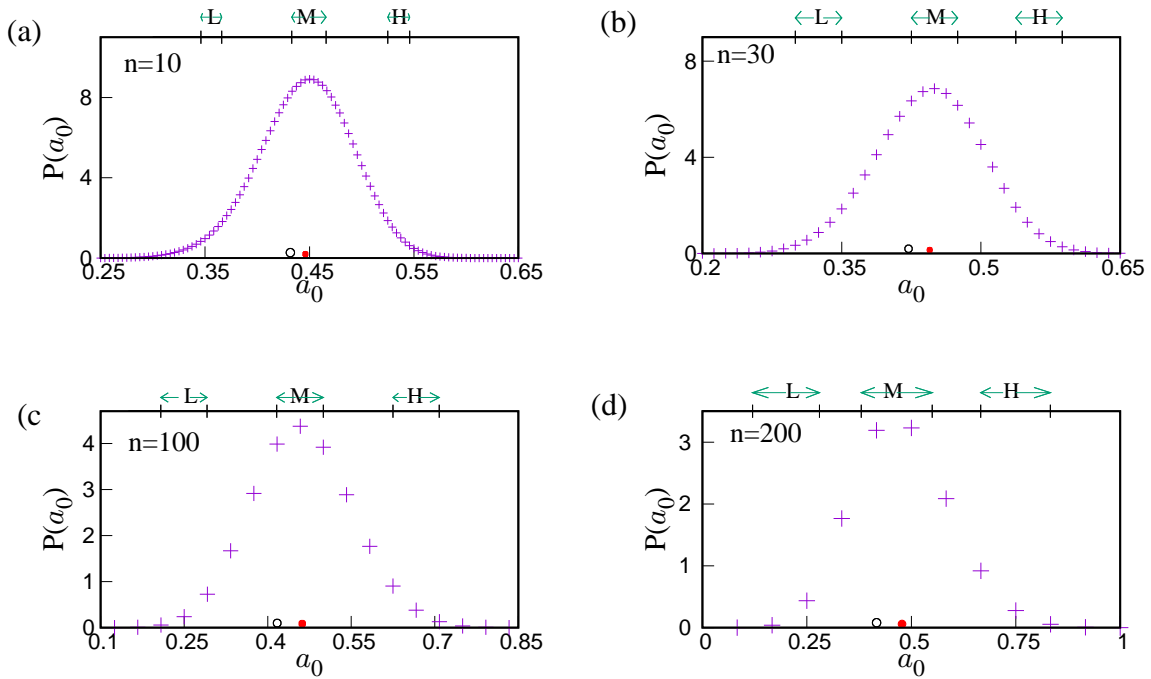}
\caption{ Distribution of activity $a_0$ at the start of a run for a strong attractant gradient.  The low, medium and high ranges of values of $a_0$ have been shown. These ranges are defined with reference to the mean $a_0$ value shown by the red point. The empty circle on the $x$-axis shows the adapted activity value which also belongs to the medium range. Each data point has been averaged over at least $10^6$ histories. All simulation parameters are listed in Table \ref{table} in Appendix \ref{app:model}. }
\label{fig:a0dists}
\end{figure}

\begin{figure}
\includegraphics[scale=0.7,angle=270]{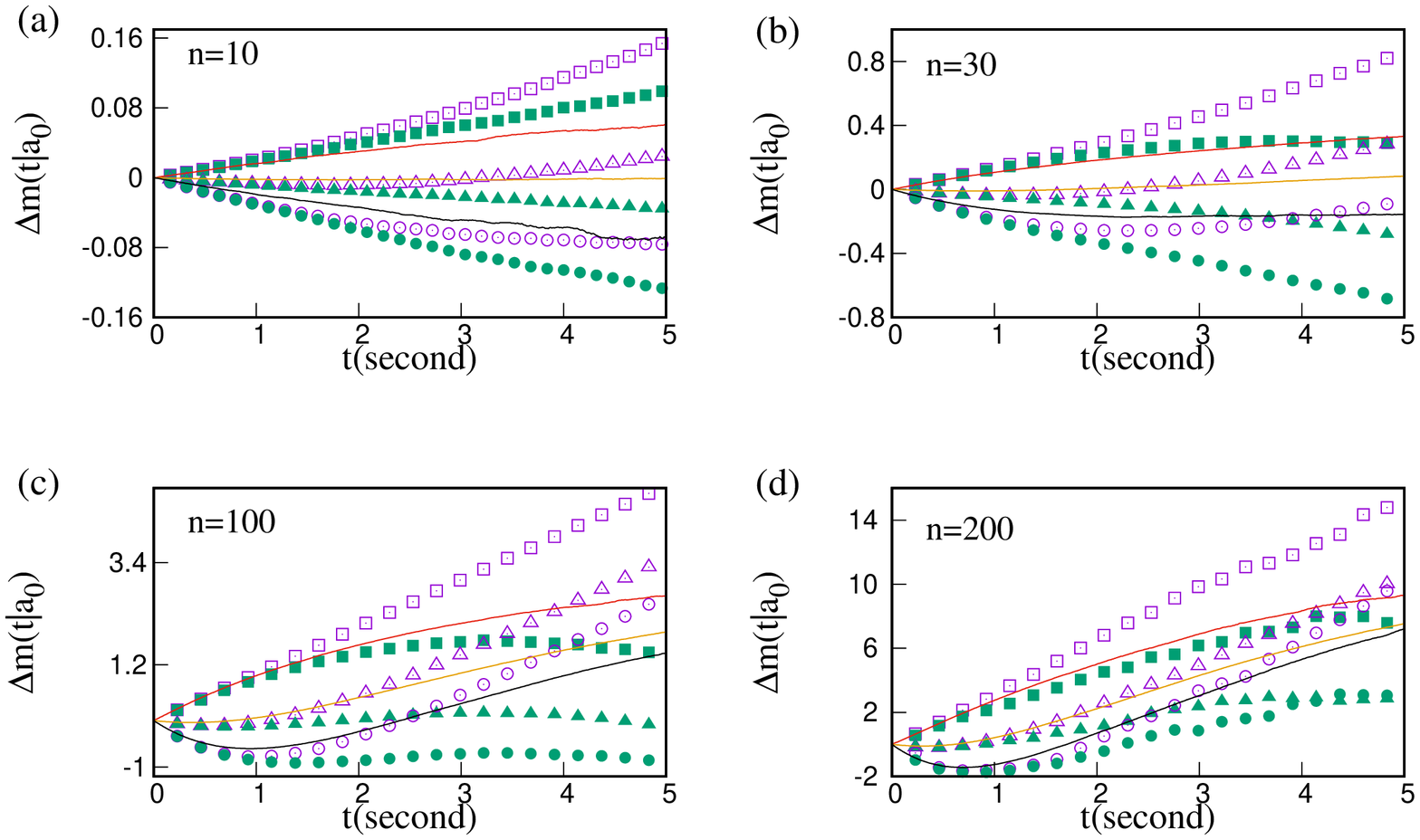}
\caption{$\Delta m^\pm (t|a_0))$ for runs starting with three different activity ranges. The discrete points present data for the strong  gradient case and the solid lines are for the flat concentration profile of the attractant. The square, triangular and circular symbols correspond to low, medium and high $a_0$ values while the empty (filled) symbols are for uphill (downhill) runs. Among the solid lines, the top (red), middle (yellow) and the bottom (black) ones correspond to low, medium and high $a_0$ runs for the flat profile. Each data point has been averaged at least $10^5$ histories. All simulation parameters are listed in Table \ref{table} in Appendix \ref{app:model}.}
\label{fig:Dmsa0}
\end{figure}

To explain this variation, we first look into the activity distribution at the start of the run and define low, medium and high ranges of activity as done in the previous subsection. In Fig. \ref{fig:a0dists} we present the data for $P(a_0)$ which is wider than the weak gradient case \cite{dufour2014limits, long2017feedback, xue2016moment, sun2017macroscopic, micali2017drift} and identify the ranges of low, medium and high $a_0$. In Fig. \ref{fig:Dmsa0} we plot $\Delta m^\pm (t|a_0))$ with $a_0$ in one of these three ranges. The solid lines in these plots show the data for the zero gradient case where we have used same ranges of $a_0$ as in Fig. \ref{fig:a0dists} which are significantly different from $a_0$ ranges for the flat attractant profile shown in Fig. \ref{fig:a0distw}. Because of this difference we sometimes find  in Fig. \ref{fig:Dmsa0}  the solid lines do not lie between the uphill and downhill curves.

 For small $n$ the behavior of $\Delta m^\pm (t|a_0))$ is qualitatively similar to the weak gradient case. For large $n$, we find all uphill runs, irrespective of their $a_0$ range, show methylation at large $t$. The uphill runs starting with large $a_0$ show initial demethylation which lower the activity. Moreover, increasing ligand concentration along the trajectory also tends to lower the activity. Due to these two effects activity falls below the adapted level and methylation takes over. For downhill runs our data in Fig. \ref{fig:Dmsa0}(b) show that low $a_0$ runs show weak methylation at large $t$ and for medium and high $a_0$ we have demethylation. In this case, the already strong gradient, further amplified by receptor cooperativity controls the behavior at large times, initial activity plays a less significant role. The fast falling ligand concentration along the trajectory raises the activity resulting in demethylation. For $n=100$ Fig. \ref{fig:Dmsa0}(c) shows that small $a_0$ runs show methylation for small times followed by a dip at larger times, medium $a_0$ runs show negligible change in methylation and high $a_0$ runs show demethylation. For $n=200$ we find medium and high $a_0$ runs show initial demethylation, followed by an upswing at late times.

Above detailed measurements help us to understand the data in Fig. \ref{fig:Dms}. Here an uphill (downhill) run experiences a steep increase (decrease) of attractant concentration with time because of the strong gradient present. For $n=10$ both $\Delta m^+(t)$ and $\Delta m^-(t)$ decrease to become negative for small $t$.  From exponential decay of $N^\pm (t)$ (see Fig. \ref{fig:runud}) it follows that the small time statistics are dominated by short runs which control the behavior of $\Delta m^\pm (t)$ for small $t$. These short runs are associated with high activity (or large tumbling bias)  which is supported by our data in Fig. \ref{fig:rundist} right panel where the high activity runs show highest probability for short durations. Such runs undergo demethylation. For moderate or large $t$, due to fast increasing ligand free energy during uphill runs, the receptor clusters tend to switch to inactive states and hence methylation takes over, making $\Delta m^+(t)$ positive again. Interestingly, in Fig. \ref{fig:Dms}(a), the red solid line that shows the methylation variation for zero gradient of the attractant, passes above the $\Delta m^+(t)$ curve at small $t$, unlike what we had seen for the weak gradient case. Although at larger $t$, the zero gradient data fall below the strong gradient data as seen in Fig. \ref{fig:Dmw} earlier, the short time behavior seems counter-intuitive. To explain this effect, we consider Fig. \ref{fig:a0distw}(a) and \ref{fig:a0dists}(a) where initial activity $a_0$ distribution at the start of a run is shown. The difference between mean $a_0$  and adapted $a_0$ is significantly larger for the strong gradient case and this makes the demethylation more pronounced. For downhill runs the short time decrease of $\Delta m^-(t)$ is expectedly more pronounced because in addition to those runs which started off with high activity, there are runs which undergo a rise in activity due to rapidly decreasing $c(x)$. The demethylation is therefore stronger in this case. As time goes on, we see $\Delta m^-(t)$ reaches a minimum and then starts increasing again. This behavior can be explained as follows. The downhill runs which persist till large times even though a strong gradient is present in the system, need to have very low activity at the start of the run and as time goes on, the fraction of such runs in $N^-(t)$ population increases with time (see data in Fig. \ref{fig:npm}(b) where the number of low activity runs overtake that of high activity runs for moderate or large $t$) . These runs undergo methylation at short times and even if their activity increases with time, leading to demethylation at large times, the net change $\Delta m^-(t)$ has a smaller magnitude. This explains the negative minimum and subsequent rise of $\Delta m^-(t)$ (also see Appendix \ref{app:pers} for supporting data). However, in the time range we have observed, $\Delta m^-(t)$ does not change its sign and continue to remain negative. For the purpose of suitable choice of scales, we have not plotted the zero gradient data with $\Delta m^-(t)$ in Fig. \ref{fig:Dms}. From the red lines in the left panels of Fig. \ref{fig:Dms} it is clear that $\Delta m^-(t)$ always lie below the zero gradient curve.

For higher values of $n$ the qualitative behavior of $\Delta m^+(t)$ remains same, but $\Delta m^-(t)$ undergoes significant change in its dynamics. For $n=30$ we find $\Delta m^-(t)$ monotonically decreases with time till the time range we have observed. In this case, the signal coming from the strong gradient gets even more amplified due to large value of $n$ and the activity state of the receptors is mainly controlled by the ligand free energy. Along the downhill trajectory of the cell, the ligand free energy decreases rapidly and this raises the activity of the receptors. Even those runs which started with a low activity value, experience an increase in activity due to this effect. Our data in Fig. \ref{fig:Dmsa0}(b) are consistent with this where we see low activity downhill runs show a decrease in methylation rate at long times. The high activity results in demethylation and  $\Delta m^-(t)$ becomes negative.

For $n=100$ the behavior is even more interesting: $\Delta m^-(t)$ reaches a negative minimum at short times, then increases to become positive, reaches a positive maximum at large times and then starts decreasing strongly again to become negative. As we explain below, this rich behavior is a result of interplay between the sensing and adaptation modules of the network \cite{shobhan}. For large receptor clusters, the activity fluctuation in the cell is quite strong. When the activity becomes too high or too low, to restore it to its average level, adaptation needs to play an important role and can sometimes override the signal coming from ligand concentration variation in the cell's environment. For small $t$, the behavior of $\Delta m^ -(t)$ is controlled by high activity runs. This can be clearly seen in Fig. \ref{fig:npm}(f) where number of high activity runs at short times is much higher than low activity ones. Note that Fig. \ref{fig:npm} show data for uphill runs but we find very similar variation for the downhill runs as well. High activity runs cause strong demethylation . However, this demethylation process lowers the activity significantly and even though ligand concentration is dropping rapidly, it cannot keep up with the strong demethylation. The resulting low activity triggers methylation (see our data in Fig. \ref{fig:Dmsa0}(c) where even for high activity downhill runs the initial demethylation slows down with time). Therefore after reaching a minimum, $\Delta m^ -(t)$ starts increasing again and even changes sign to become positive. At large times, when only very long runs survive in $N^-(t)$ population, the drop in ligand free energy along such long downhill trajectories becomes quite large which can now compete against the strong methylation variation experienced by the receptors. Therefore, the activity starts increasing again which explains the maximum of $\Delta m^ -(t)$ followed by a drop (our data for low activity downhill runs in Fig. \ref{fig:Dmsa0}(c) show this trend clearly). We have presented additional supporting data in Appendix \ref{app:pers}.

In Fig. \ref{fig:Dms} we also show data for $n=200$ where the pattern of variation is almost similar for both $\Delta m^+(t)$ and $\Delta m^-(t)$, except at very large times when $\Delta m^-(t)$ shows a flattening tendency. The similarity between the uphill and downhill runs shows that the methylation dynamics is insensitive to the ligand density. In this range of $n$ values, the adaptation module wins over the sensing module of the signaling network because of large activity fluctuations. Irrespective of whether the cell is headed uphill or downhill along the ligand concentration profile, the activity of the receptors remains low in the run state resulting in overall methylation. The late time flattening tendency of $\Delta m^-(t)$ is nothing but a remnant of the behavior seen at $n=100$ where very large drop in ligand density during very long downhill runs finally tends to increase the activity. 

Our data for $\delta m^\pm (t)$ for the strong gradient case has been presented in Appendix \ref{app:dms1}. A more detailed quantity which tracks the distribution of the methylation level with time $t$ has been shown in Fig. \ref{fig:mtdist} of Appendix \ref{app:mdis} and is found to be consistent with the variation of $\delta m^\pm (t)$.

\section{Conclusions}
\label{sec:con}
In this study we have performed a detailed analysis of methylation dynamics of chemoreceptors of an {\sl E.coli} cell while the cell is swimming in a spatially varying  attractant environment. We have considered  attractant concentration with strong and weak spatial gradient. Our numerical simulations show that the receptor cooperativity strongly affects the methylation dynamics and the effect is rather dramatic in the case of strong gradient of the attractant. In all cases we find that at short times the methylation dynamics is controlled by short runs which are generally associated with high activity. This causes the average methylation of the receptors decrease with time initially after the start of a run. For intermediate or large times, the methylation dynamics is controlled by long runs and depending on the receptor cooperativity or strength of the gradient, the qualitative behavior of methylation can be completely different in this time regime. For weak gradient during both uphill and downhill runs, after the initial demethylation, we find methylation level increases steadily with time, till the time range we have observed. Although the quantitative range of variation increases with size of the receptor clusters, the qualitative behavior remains the same. Similar temporal variation is observed for uphill runs in the strong gradient case as well. But for downhill runs with strong gradient, the methylation dynamics shows highly non-trivial dependence on receptor cooperativity. For relatively small size of the receptor clusters, the long downhill runs must start with low activity and hence they undergo methylation at short times. This early methylation combined with decreasing ligand concentration with time, raises the activity and hence demethylation happens at large times. As the receptor cluster size increases, the input signal coming from ligand concentration gets amplified and gains control over the activity variation. For downhill runs this causes increase in activity, even for those long runs whose activity was low at the beginning. Therefore, methylation level monotonically decreases with time in this case. As the cluster size increases further, adaptation tends to win over sensing. The cell is less sensitive to ligand concentration variation now: only when a downhill run has persisted for a really long time during which the drop in ligand free energy has been quite large, the activity gets controlled by ligand concentration. For even larger receptor clusters, adaptation wins over sensing at all times and the methylation dynamics for uphill and downhill runs start looking almost similar.

To the best of our knowledge, such a systematic, quantitative investigation of methylation dynamics has never been performed before, even though {\sl E.coli} chemotaxis is a widely studied system. Our study provides probing insights into how the amplification of input signal and a negative feedback mechanism come together to control the time-evolution of various dynamical variables which characterize the signaling network. We find highly interesting and non-trivial methylation dynamics as a result of this interplay. Our results can be tested in experiments. Although our study focuses on a swimming cell, our conclusions can be tested for a tethered cell which is experimentally more accessible.  In \cite{shimizu2010modular} the real time activity of the receptor-kinase complex was measured using FRET and by subjecting the tethered cell to time-varying ligand concentration such that the FRET output remains constant in time, the adaptation dynamics was determined.   Motivated by this,  we propose an experiment on a tethered cell with appropriately engineered attractant environment to verify our conclusions. The counter clockwise rotation of the flagellar motors of the tethered cell can be considered equivalent to a run, and by ramping up (down) the attractant level at the fixed location of the tethered cell one can mimic an uphill (a downhill) run. We have included a detailed discussion of the experimental protocol in Appendix \ref{app:expt}. The activity level of the tethered cell can be tracked during the changing attractant concentration from which the temporal variation of methylation level can be determined. Possible future experimental developments that enable {\sl in vivo} measurement of methylation levels directly, can also  be useful to test our conclusions.

There are few aspects of the signaling network, which we have not taken into account in our model. The receptor arrays are known to show a hexagonal symmetry in their spatial arrangement \cite{briegel2012bacterial, liu2012molecular} which we have not considered in our model. Moreover, it has been observed that receptor clusters cause curvature of the cell membrane which has an energy cost \cite{endres2009polar, haselwandter2014role, draper2017origins}. We do not include this effect in our simple model. These assumptions may make quantitative comparison between our results and experiments difficult, but they are unlikely to affect our general conclusions. The understanding of the methylation dynamics that our study provides, is much more general and simply relies on the coupling between ligand concentration, activity and methylation, and does not depend on the details of the model. It will be of interest to see if our understanding applies to other sensory systems as well.

\section{Acknowledgements} 
SDM acknowledges a research fellowship [Grant No. 09/575(0122)/2019-EMR -I]  from the Council of Scientific and Industrial Research (CSIR), India. SC acknowledges financial support from the Science and Engineering Research Board, India (Grant No: MTR/2019/000946).

\appendix
\newpage
\section{Additional details of the model}
\label{app:model}
\renewcommand{\thefigure}{A\arabic{figure}}
\setcounter{figure}{0}

In our model, there are three major parts: (a) activity switching of the receptor clusters, (b) binding-unbinding dynamics of the enzyme molecules to the receptor dimers and (de)methylation of the receptor dimers by the bound enzymes, (c) run-and-tumble motion of the cell where the switching probability between the run mode and tumble mode is calculated from the total activity. A detailed description of each part follows below.

(a) We denote the activity state of the $i-$th receptor cluster containing $n$ trimers of dimers by the variable $a_i$, which can take two values. $a_i=1$ denotes an active state and $a_i=0$ denotes inactive state. The free energy difference $F$ between these two states is given by Eq. \ref{eq:fcl} of the main paper and the probability to find a receptor cluster in active state is $[1+\exp(F)]^{-1}$. From $a_i=0$ state the receptor cluster switches to $a_i=1$ state with the rate $\dfrac{w_a}{1+\exp(F)}$ and the reverse transition happens with a rate $\dfrac{w_a \exp(F)}{1+\exp(F)}$ \cite{colin2017multiple}. The choice of these rates is based on local detailed balance \cite{colin2017multiple}. The parameter $w_a$ is the characteristic time-scale of the transition \cite{shobhanijp}.

(b) The total number of CheR and CheB molecules are denoted by $N_R$ and $N_B$, respectively. An unbound CheR molecule resides in the cell cytoplasm and can bind to the receptor dimer if an only if no other enzyme is bound to it. The binding can take place at the tether site or the modification site of the receptor dimer\cite{pontius2013adaptation, wu1996receptor, feng1999enhanced}. While both these bindings are slow, the tether binding is comparatively faster than the binding at the modification site \cite{schulmeister2008protein, pontius2013adaptation} because of which we only consider tether binding process in our model. The binding takes place with rate $w_r$ and once bound the CheR enzyme raises the methylation level of the dimer by one unit with rate $k_r$, provided the dimer belongs to an inactive cluster and its methylation level is $< 8$. A bound CheR can unbind from the dimer with rate $w_u$ and can either reattach to another unoccupied dimer within the same cluster, or return to the cytoplasm. A CheB molecule in the cytoplasm can undergo phosphorylation by an active receptor with rate $w_p$ and an unbound CheB-P molecule can dephosphorylate with rate $w_{dp}$. The binding, rebinding and unbinding processes for CheB-P are very similar to those for CheR, while demethylation happens if the dimer is active with a non-zero methylation level. The binding between an enzyme molecule and a dimer is a slow process and therefore if one binding event results in only one (de)methylation reaction, then it becomes difficult for the network to maintain perfect adaptation. The possibility of rebinding of the same enzyme molecule to another dimer in the same cluster paves way for multiple (de)methylation of multiple receptors from a single binding event. This is an effective way to include the flavor of  assistance neighborhood \cite{endres2006precise, hansen2008chemotaxis, li2005adaptational} and brachiation \cite{levin2002binding} in our model.

(c) Let $a$ denote the fraction of active receptor clusters in the cell. Then the phosphorylated fraction of CheY molecules, defined as $Y_P = \dfrac{[\text{CheY-P}]}{[\text{CheY}]}$ follows the rate equation \cite{flores2012signaling}
\be 
\frac{dY_P}{dt} = K_Y a (1-Y_P) - K_Z Y_P
\ee
where the parameters $K_Y$ and $K_Z$ are rates for phosphorylation and dephosphorylation, respectively. The run-tumble motility of the cell is controlled by $Y_P$. If the cell is in the run mode, it can switch to tumble mode with rate $\omega \exp (-G)$ where $G = \Delta_1 - \dfrac{\Delta_2}{1+Y_0/Y_P}$ and the opposite switch from tumble to run happens with the rate $\omega \exp(G)$. We have verified (data not shown here) that our conclusions remain unaffected even if tumble to run switch is assumed to be independent of $Y_P$ and takes a constant value.

We present the values of all model parameters in Table \ref{table}. 

\begin{center}
\begin{table}[h]
\caption{}
\begin{tabular}{|l|l|l|l|}
\hline
\textbf{Symbol} & \hspace{30mm} \textbf{Description} & \textbf{Value} & \textbf{References} \\
\hline
$N_{dim}$ & Total number of receptor dimers & $7200$ & \cite{pontius2013adaptation,li2004cellular} \\
\hline
$N_R$  & Total number of CheR  protein molecules &  $140$ & \cite{pontius2013adaptation,li2004cellular}\\
\hline
$N_B$  & Total number of CheB  protein molecules &  $240$ & \cite{pontius2013adaptation,li2004cellular} \\
\hline
 $\epsilon_0$ &  Basal energy of receptor dimer &  $1$ $k_B T$   &  \cite{pontius2013adaptation,frankel2014adaptability,dufour2014limits,long2017feedback} \\
\hline
 $\epsilon_1$ &  Receptor energy change per methyl group addition &  $1$ $k_B T$  &  \cite{pontius2013adaptation,frankel2014adaptability,dufour2014limits,long2017feedback}  \\
\hline
$K_{min}$ & Minimum concentration receptor can sense &  $18$ $\mu M$ & \cite{jiang2010quantitative}, \cite{flores2012signaling} \\
\hline
$K_{max}$ & Maximum concentration receptor can sense &  $3000$ $\mu M$ & \cite{flores2012signaling,jiang2010quantitative} \\
\hline
$w_{a}$  & Flipping rate of activity &  $0.75$ $s^{-1}$ & Present study\\
\hline
$\omega$  & Switching frequency of motor &  $1.3$ $s^{-1}$ & \cite{sneddon2012stochastic, sneddon2011efficient} \\
\hline
 $\Delta_{1}$  & Nondimensional constant regulating motor switching  &  $10$ & \cite{sneddon2012stochastic, sneddon2011efficient} \\
\hline
$\Delta_{2}$  & Nondimensional constant regulating motor switching &  $20$ & \cite{sneddon2012stochastic, sneddon2011efficient}  \\
\hline
$Y_{0}$ & Adopted value of the fraction of CheY-P protein &  $0.34$ & \cite{sneddon2012stochastic, sneddon2011efficient} \\
\hline 
$K_{Y}$ & Phosphorylation rate of CheY molecule &  $1.7$ $s^{-1}$ & \cite{flores2012signaling,tu2008modeling} \\
\hline
$K_{Z}$ & Dephosphorylation rate of CheY molecule &  $2$ $s^{-1}$ & \cite{flores2012signaling,tu2008modeling}\\
\hline
$w_{r}$ & Binding rate of bulk CheR to tether site of an unoccupied dimer &  $0.068$ $s^{-1}$ & \cite{pontius2013adaptation,schulmeister2008protein} \\
\hline
$w_{b}$ & Binding rate of bulk CheB-P to tether site of an unoccupied dimer &  $0.061$ $s^{-1}$ & \cite{pontius2013adaptation,schulmeister2008protein} \\
\hline
$w_{u}$ & Unbinding rate of bound CheR and CheB-P &  $5$ $s^{-1}$ & \cite{pontius2013adaptation,schulmeister2008protein} \\	
\hline	
$k_{r}$ & Methylation rate of bound CheR & $2.7$ $s^{-1}$ & \cite{pontius2013adaptation,schulmeister2008protein}\\ 	
\hline
$k_{b}$ & Demethylation rate of bound CheB-P &  $3$ $s^{-1}$ & \cite{pontius2013adaptation,schulmeister2008protein}\\
\hline	
$w_{p}$ & CheB phosphorylation rate &  $3$ $s^{-1}$ & \cite{pontius2013adaptation,stewart2000rapid} \\
\hline
$w_{dp}$ & CheB-P dephosphorylation rate &  $0.37$ $s^{-1}$ & \cite{pontius2013adaptation}\\	
\hline	
$L$ &  Box length in 1D &  $2000$ $\mu m$ & Present study \\
\hline
$v$ & Speed of the cell &  $20$ $\mu m/s$ & \cite{berg2008coli} \\
\hline
$dt$ & Time step &  $0.01$ $s$ & Present study\\
\hline
$L_{x}\times L_{y}$ & Box dimension in 2D&  $2000\times 800$ $\mu m^{2}$ & Present study \\
\hline
$D_{\Theta}$ & Rotational Diffusivity &  $0.062$ $\mu m^{2}/s$ & \cite{berg1972chemotaxis,dufour2014limits,karmakar2016enhancement}\\
\hline
$c_{0}$ & Background attractant concentration &  $200$ $\mu M$ & Present study \\
\hline
$1/x_{0}$ & Linear concentration gradient of  attractant &  $0.025$ $ mm^{-1}$, $0.25$ $m m^{-1}$ & Present study \\
\hline
\end{tabular}
\label{table}
\end{table}
\end{center}

\newpage
\section{Simulation details and suggested experiments}
\label{app:expt} \renewcommand{\thefigure}{B\arabic{figure}}  \setcounter{figure}{0}

In our agent-based simulations, we consider run-and-tumble motion of the cell in $1d$ and $2d$ box with reflecting boundary walls. A linear concentration profile for the  attractant is set up along the $x$-direction. During a run the cell moves with a constant speed $v$. After each tumble, the direction of the new run is chosen at random. In $1d$ case, the trajectory during a run is a straight line, while in $2d$ case due to rotational diffusion the trajectory shows gradual bending. After each tumble the cell chooses a random direction to start a new run. We perform all measurements in the long time limit when the system has reached a steady state. We average over all runs that originate beyond a distance from the boundary walls. This ensures negligible boundary effect. In our simulations we use $x_d=400 \mu m$ and $y_d=200 \mu m$ as the width of the boundary layers in $x$ and $y$ directions, respectively. We have verified (data not shown here) that our results are not sensitive to the choice of boundary layer width.

We use a  attractant concentration profile $c(x)=c_0(1+x/x_0)$. In the $1d$ case when the cell runs rightward, it experiences a linear increase of level with time along its trajectory. Similarly, during a leftward run, the  attractant level drops linearly with time. At $x=0,L$ there are reflecting boundary walls and when a cell hits these walls they reverse their run directions. To recreate these conditions in experiment with a tethered cell, one needs to use a suitably engineered attractant environment. The counter clockwise rotation of the flagellar motors of the tethered cell can be considered equivalent to a run. An uphill (downhill) run can be mimicked by ramping up (down) the attractant level linearly with time at the fixed location of the tethered cell. The ramping rate can be chosen to be exactly same as the rate at which a cell running with speed $v$ experiences change in attractant levels along its path. Of course for weak and strong gradient, this rate is going to be different. Every time the flagellar motors switch to a clockwise rotation, the attractant level should be held fixed as this corresponds to a tumble mode with zero displacement. After each tumble, when the motors switch back to counter clockwise rotation, the sign of the ramp rate can be chosen at random. Finally, when the attractant level matches the boundary values, the ramp rate should simply be reversed, which corresponds to the cell hitting a boundary wall and getting reflected back and continuing its run in the opposite direction. This way by tuning the ramp rate in sync with the rotational bias of the flagellar motors, we can create the same conditions of a swimming cell for a tethered one. By tracking the methylation level of the receptor clusters for this tethered cell during the ramped up and ramped down attractant level, one can directly measure the quantities like $\Delta m^\pm (t)$ or $\delta m^\pm (t)$ and test our conclusions. Alternatively, the activity level of the cell can be tracked and the methylation variation can be determined from there by using the knowledge of input variation of ligand concentration.

\newpage
\section{Exponential decay of $N^\pm(t)$} 
\renewcommand{\thefigure}{C\arabic{figure}}  \setcounter{figure}{0}
\begin{figure}[h]
\includegraphics[scale=1.5]{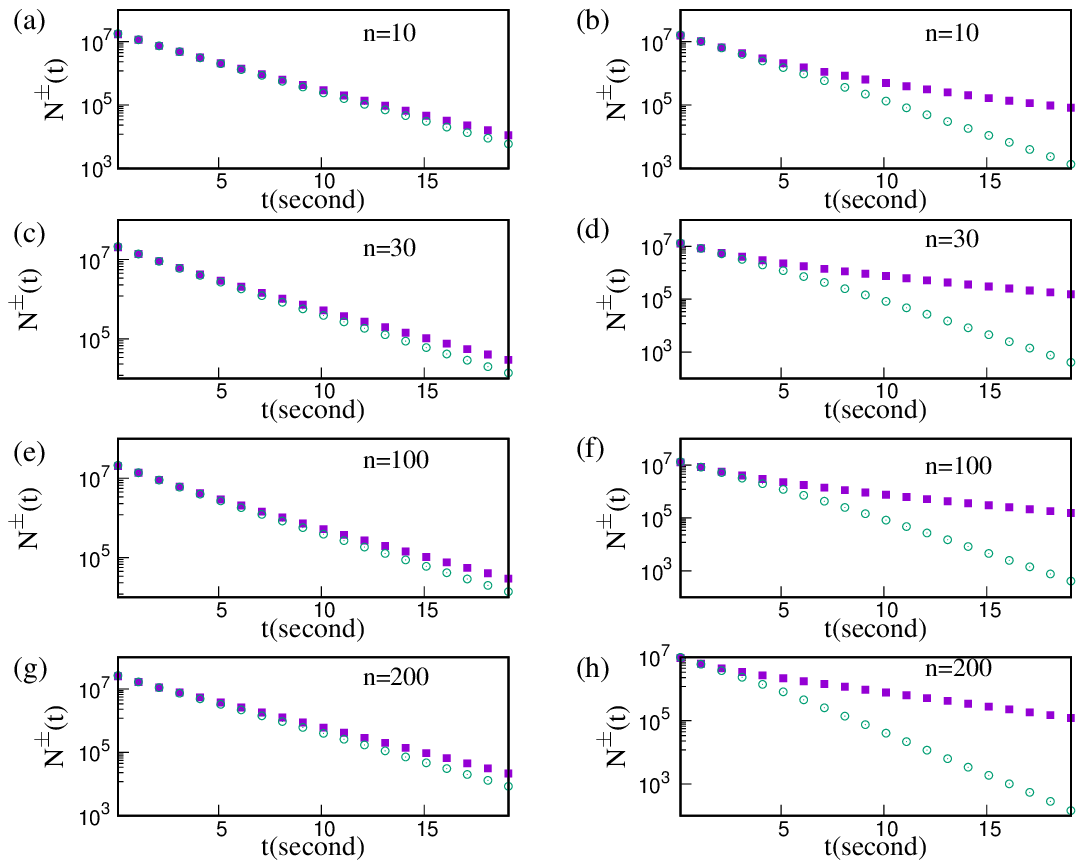}
\caption{ Number of surviving runs as a function of time for both uphill(purple solid square) and downhill(green empty circles) runs. Left panel is for weak gradient and right one is for strong gradient. It shows that for both weak and strong gradient number of downhill runs drops faster than number of uphill ones. Each data has been averaged over at least $10^5$ histories. All simulation parameters are listed in Table \ref{table} in Appendix \ref{app:model}}
\label{fig:runud}
\end{figure}

\newpage
\section{Runs starting with low, medium and high activity }
\label{app:lowhigh}  
\renewcommand{\thefigure}{D\arabic{figure}}  \setcounter{figure}{0}

\begin{figure}[h]
\includegraphics[scale=1.5]{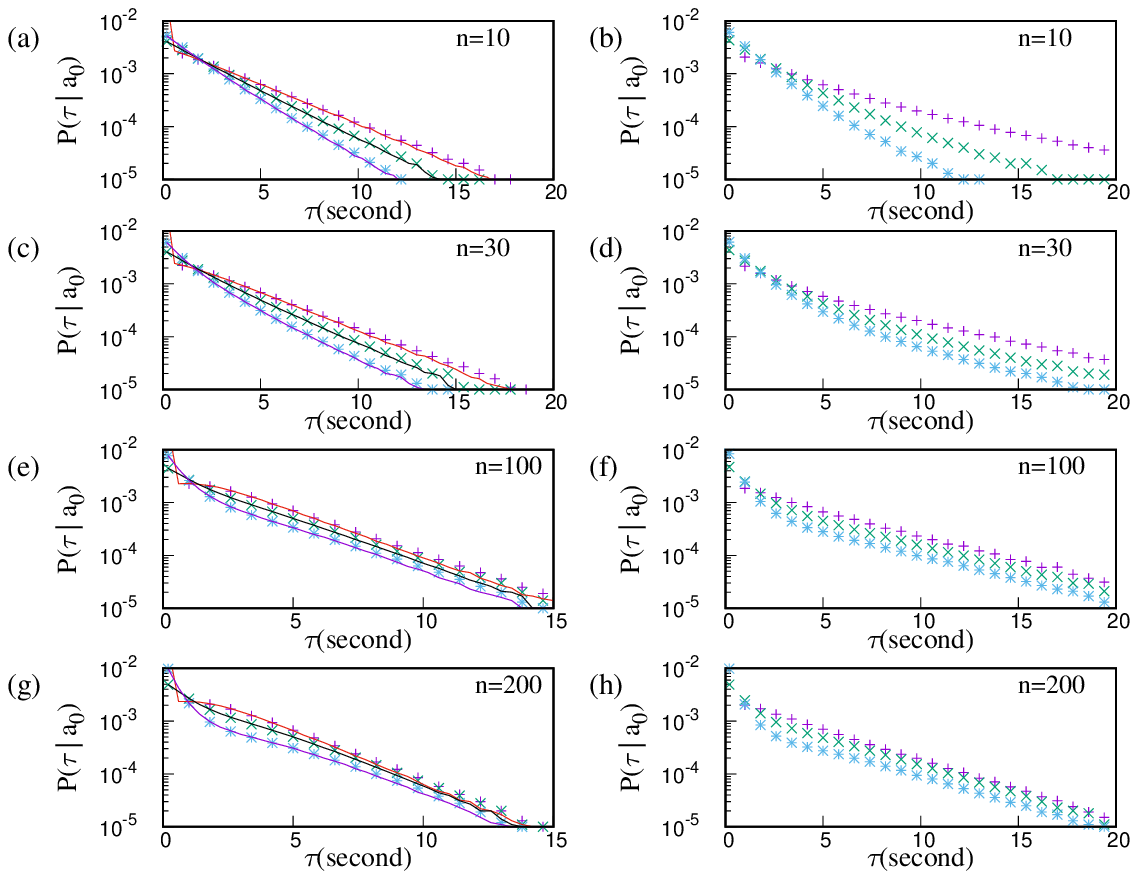}
\caption{ Distribution of run durations starting with low (purple plus), medium (green cross) and high (blue star) values of initial activity. The left panel shows data for weak gradient and the right panel is for stronger gradient.  The zero gradient data have been shown by continuous lines in the left panel. The purple, green and blue lines correspond to low, medium and high $a_0$, respectively. As expected, runs starting with lower activity values survive the longest. Each data point has been averaged over at least $2\times 10^6$ times. All simulation parameters are listed in Table \ref{table} in Appendix \ref{app:model}}
\label{fig:rundist}
\end{figure}

\begin{figure}[h]
\includegraphics[scale=1.5]{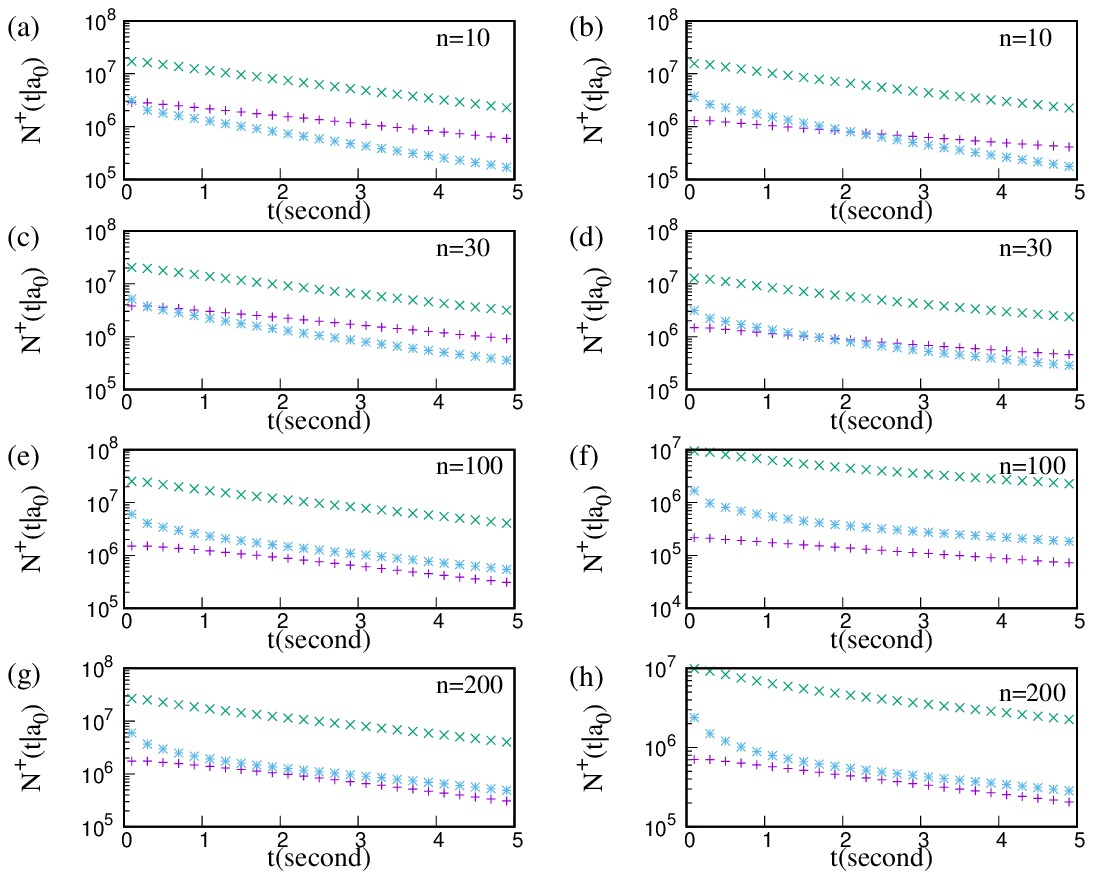}
\caption{ Number of surviving runs as a function of time for runs starting with three activity zones, which are  high activity (blue star), medium activity (green cross) and low activity (purple plus). Left (right) panel is for weak (strong) gradient. Each data point has been averaged over at least $10^5$ histories. All simulation parameters are listed in Table \ref{table} in Appendix \ref{app:model}.}
\label{fig:npm}
\end{figure}

Fig. \ref{fig:rundist} shows distribution for duration of runs which start with activity values in high, medium or low ranges. To measure the persistence of these runs, in Fig. \ref{fig:npm} we plot $N^+(t|a_0)$, the number of uphill runs which do not tumble till time $t$. As expected, for all gradient strength and all cooperativity, $N^+(t|a_0)$ is largest for runs starting with medium range of $a_0$ (green filled circles) because that is where the most probable value of $a_0$ lies (shown by red dot in each panel of Fig. \ref{fig:a0distw} and \ref{fig:a0dists}). Fig. \ref{fig:npm} also shows that $N^+(t|a_0)$ has sharpest drop for high $a_0$ (blue empty triangles), specially at small $t$ since these are the runs associated with high tumbling bias. We find very similar behavior for $N^-(t|a_0)$ also (data not shown here).

\newpage
\section{Average activity in uphill and downhill runs}	
\label{app:act}
\renewcommand{\thefigure}{E\arabic{figure}}  \setcounter{figure}{0}
Let $a^+$ denote the activity of the cell measured at a random time during an uphill run. Clearly, $a^+$ is a stochastic quantity and we denote its mean by $\langle a^+ \rangle$. Similarly, for downhill runs $\langle a^- \rangle$ can be defined. Our data in Fig. \ref{fig:act} show that the difference between these two activities decreases with $n$ for large $n$. This is consistent with the fact that for large $n$ adaptation wins over sensing and the cell is less sensitive to ligand concentration variation in its surroundings. 
\begin{figure}[h]
\includegraphics[scale=0.4,angle=270]{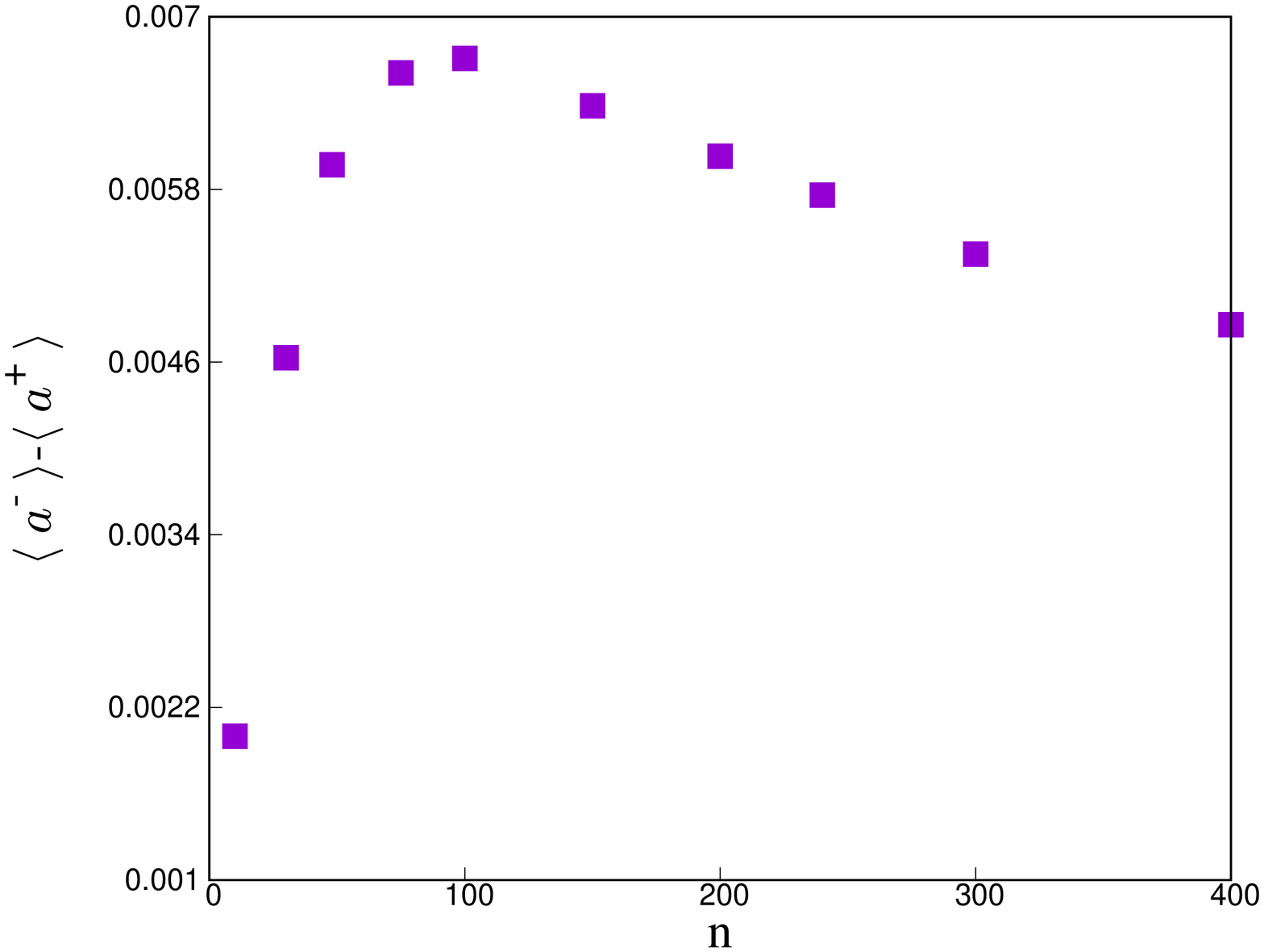}
\caption{Difference of average activity during uphill and downhill runs. The difference decreases with $n$ for large $n$ which reflects dominance of adaptation module over sensing module. These data are for $1d$ motion of the cell with weak gradient of $c(x)$. Each data point has been averaged over at least $10^7$ histories. All simulation parameters are listed in Table \ref{table} in Appendix \ref{app:model}}
\label{fig:act}
\end{figure}

\newpage
\section{Methylation dynamics for very long runs}
\label{app:pers}
\renewcommand{\thefigure}{F\arabic{figure}}  \setcounter{figure}{0}
The time evolution of methylation level during particularly long runs provides an independent verification of the explanation we have provided for the behavior of $\Delta m^-(t)$ in Fig. \ref{fig:Dms}. To this end we perform the following measurement. Let ${\cal M}^-(t,\tau)$ be the methylation level of a receptor cluster at time $t$ during a downhill run which persists for at least time $\tau$. In Fig. \ref{fig:pers} we present data for average change in methylation (average calculated over $N^-(\tau)$ runs) as a function of time $t$ for $\tau = 5s$ which is much longer than average run duration. These data are for the strong gradient case. For $n=10$ a long run starts with low activity and hence for short times there is methylation. But decreasing $c(x)$ along with increasing methylation level finally raise the activity and demethylation starts at large times. Our data in Fig. \ref{fig:Dms} top right panel show that for large $t$ when long runs dominate, the magnitude of $\Delta m^-(t)$ decreases with time. This is consistent with our data in Fig. \ref{fig:pers} top left panel where because of initial methylation and subsequent demethylation, the net change in methylation level becomes small. The top right panel in Fig. \ref{fig:pers} shows data for $n=30$ where a monotonic decrease is observed. Because of stronger receptor cooperativity in this case, the input signal coming from $c(x)$ is significantly amplified and dominates the free energy in Eq. \ref{eq:fcl}. So even if the long runs started with low activity, under the influence of rapidly decreasing $c(x)$, activity increases and as a result demethylation happens. This is consistent with our data for $n=30$ in Fig. \ref{fig:Dms}. For $n=100$ the long runs show even more interesting behavior. Adaptation plays important role here and the free energy in Eq. \ref{eq:fcl} is not controlled by ligand density alone, methylation starts playing a more important role. As we see from the bottom left panel of Fig. \ref{fig:pers}, after initial demethylation, the activity drops significantly and methylation starts. In this time regime methylation dominates over ligand free energy. But at late times, when the downhill run has gone on for quite long, the drop in $c(x)$ becomes so large that activity is now controlled by ligand density and demethylation happens again. This behavior mirrors what we had seen for $\Delta m^-(t)$ in Fig. \ref{fig:Dms} for $n=100$. The bottom right panel of Fig. \ref{fig:pers} shows the data for $n=200$ where adaptation wins over sensing at all times, and even through $c(x)$ is decreasing along the cell trajectory, that is not enough to raise the activity. We find activity remains low and methylation happens at all times during the long runs. 
\begin{figure}[h]
\includegraphics[scale=0.5,angle=270]{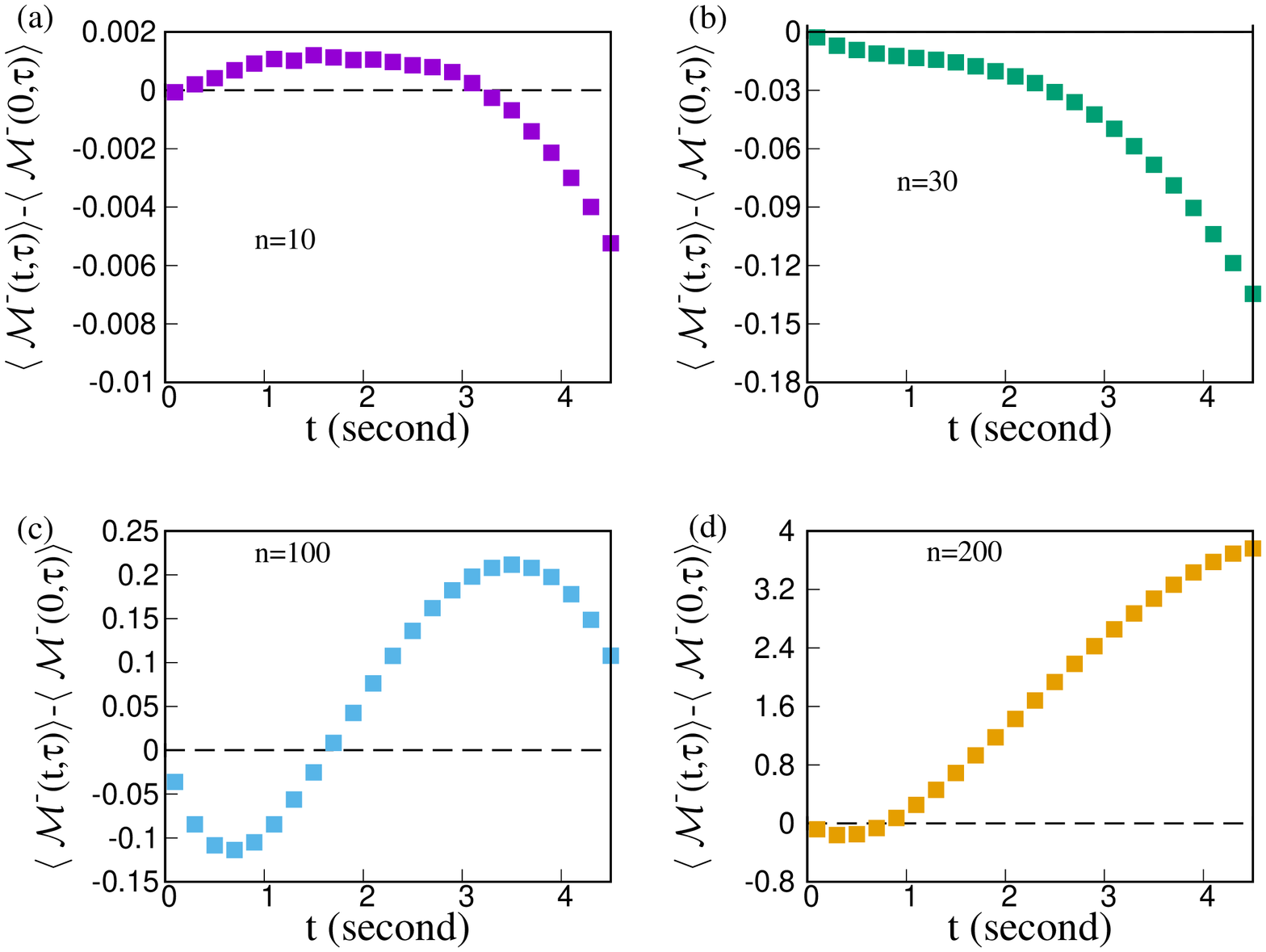}
\caption{Average change in methylation for long downhill runs with duration longer than $5s$. Here, strong gradient of $c(x)$ is used. For different values of $n$ these data correctly reflect the long time behavior of $\Delta m^-(t)$ shown in Fig. \ref{fig:Dms}. These data have been averaged over at least $10^6$ histories. Other simulation parameters are as in Fig. \ref{fig:Dms}.}
\label{fig:pers}
\end{figure}

\newpage 
\section{$\delta m^\pm (t)$ for strong gradient in one dimension }
\label{app:dms1} 
\renewcommand{\thefigure}{G\arabic{figure}} \setcounter{figure}{0}
In Fig. \ref{fig:dms} we present data for $\delta m^\pm (t)$ (purple squares). As seen in the case of weak gradient, here also we find qualitatively different trend from $\Delta m^\pm (t)$ (see Fig. \ref{fig:Dms}). For small $n$, when the variation of $\Delta m^\pm (t)$ is relatively milder, due to systematic dropping out of high methylation states from $N^\pm (t)$ populations, we find $\delta m^\pm (t)$ decrease monotonically. For $n=30$, we have already shown from our data in Fig. \ref{fig:Dms} that $\Delta m^- (t)$ shows monotonic decrease with $t$. With dropping out of high methylation states with time, the decrease is now (quantitatively) stronger for $\delta m^ - (t)$. For the uphill runs, $\delta m^+(t)$ remains negative but shows a minimum. The strong positive growth of methylation level of individual trajectories, as captured by $\Delta m^+(t)$, together with termination of high methylation runs with time, lowers the magnitude of $\delta m^+(t)$ at large times. Similar explanation can be used to interpret the data for $\delta m^+(t)$ for $n=100,200$ as well. However, $\delta m^-(t)$ for $n=200$ shows a different behavior. After an initial minimum it shows a shallow maximum followed by a steep drop. While the explanation for the initial minimum remains same as in $\delta m^+(t)$ case, the late time steep drop can be traced back to the slower growth of $\Delta m^-(t)$ at late times in  Fig. \ref{fig:Dms}h. This slower growth combined with high drop out rate of high $m$ runs from $N^-(t)$ population causes the sharp decline in $\delta m^-(t)$ at large times.
\begin{figure}[h]
\includegraphics[scale=1.2]{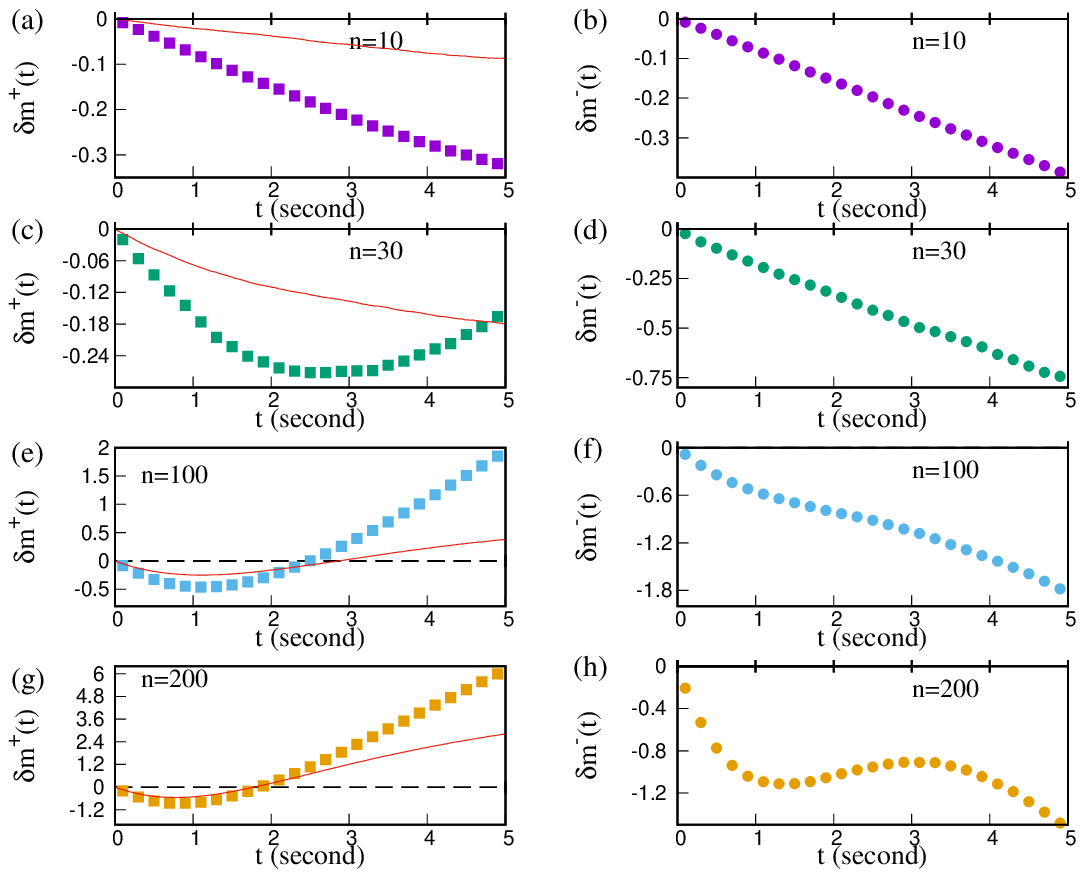}
\caption{Temporal variation of $\delta m^\pm (t)$ for different $n$ and strong gradient: left (right) panel shows data for uphill (downhill) runs. While for small $n$ both uphill and downhill runs show decreasing level of average methylation with time, there is a trend reversal for uphill runs at late times for larger values of $n$. This is caused by strong positive growth of $\Delta m^+(t)$ which overcompensates for dropping out of large methylation states from $N^+(t)$ population. These data have been averaged over at least $10^6$ histories. Other simulation parameters are as in Fig. \ref{fig:Dms}.}
\label{fig:dms}
\end{figure}

The red lines in Fig. \ref{fig:dms} left panels show data for the zero gradient case. Since the attractant concentration does not change with time in this case, one would expect less pronounced demethylation than an uphill run in strong gradient where the attractant concentration increases with time. However, our data in Fig. \ref{fig:dms}(a) show that the zero gradient data lie above the strong gradient uphill data. The explanation of this effect can be found in the distribution of initial activity $a_0$ in Figs. \ref{fig:a0distw} and \ref{fig:a0dists} where the difference between mean $a_0$ and the adapted activity is larger for the case of strong gradient, which gives rise to stronger demethylation. From Fig. \ref{fig:a0dists} it also follows that for larger $n$ values the difference between mean $a_0$ and adapted activity is comparable for the zero gradient and strong gradient case. Moreover, the increase in ligand free energy with time along an uphill run is significantly higher for large $n$ and strong gradient. This is why the zero gradient data fall below $\delta m^+(t)$ for large $n$ and large $t$.

\newpage
\section{Distribution of methylation level at time $t$ during a run}
\label{app:mdis}
\renewcommand{\thefigure}{H\arabic{figure}} 
\setcounter{figure}{0}
Let $m_t$ be the methylation level of a receptor cluster (rescaled by the size of the cluster) at time $t$ during a run. At $t=0$ we have $m_0$ that denotes the initial methylation level at the start of a run. In Fig. \ref{fig:mtdist} left and middle panels we show the distribution of $m_t$ for $t=0,2,5 $ seconds for the strong gradient case. The left panel shows the data for the uphill runs and the middle panel corresponds to downhill runs. The right panel in this figure show data for a flat attractant profile. In this case for $n=10$ and $30$ the methylation distribution shows a distinct second peak. We have not been able to explain this effect. However, the time-dependence of these curves are exactly as one would expect from $\delta m^\pm (t)$ variation shown in Fig. \ref{fig:dms}. In those cases when $\delta m^\pm (t)$ decreases monotonically with $t$, we find the distribution $P(m_t)$ also shifts leftward towards smaller $m_t$ values as $t$ increases. On the other hand, in Figs. \ref{fig:dms}e or   \ref{fig:dms}g, where $\delta m^+(t)$ shows distinct non-monotonicity along with zero-crossing and change of sign, corresponding $P(m_t)$ also shows analogous behavior. 
\begin{figure}[h!]
\includegraphics[scale=1.5]{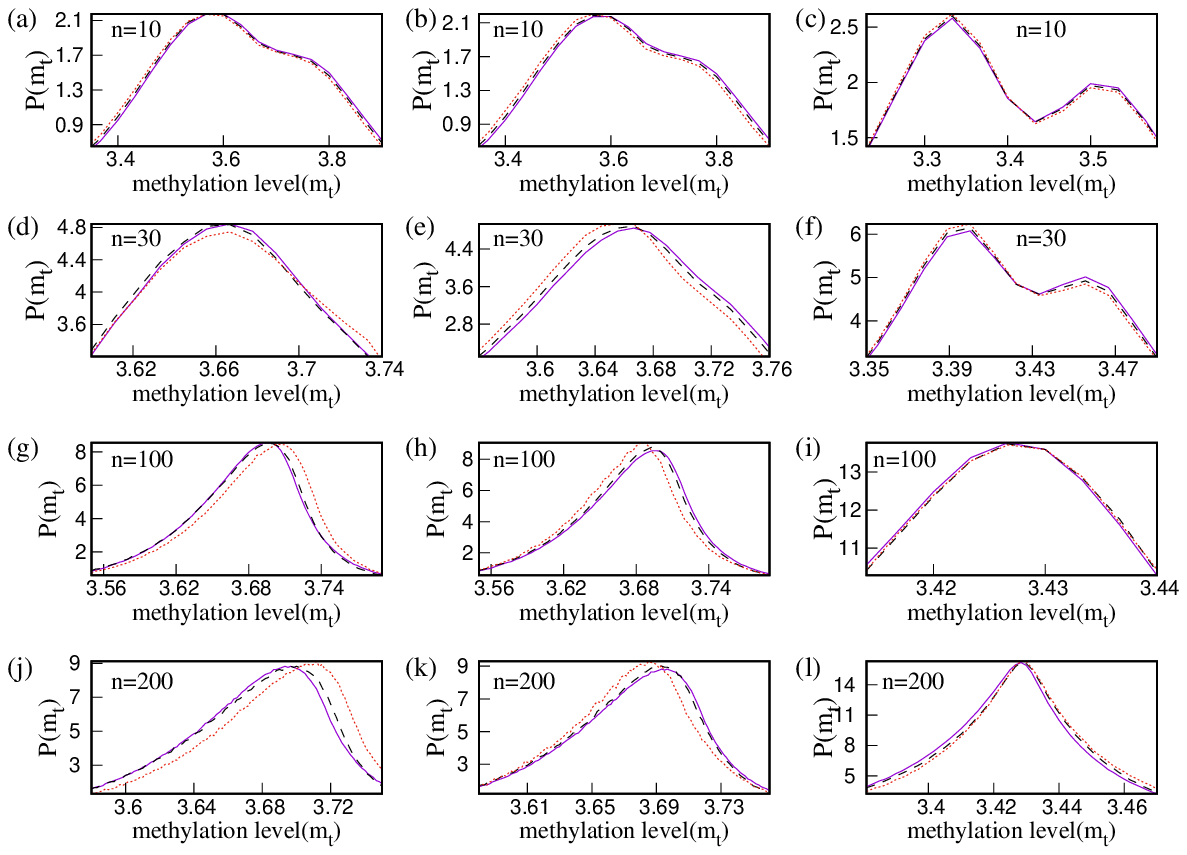}
\caption{ Distribution of methylation level at $t=0$ (solid purple line), $t=2s$ (dashed black line) and $t=5s$ (dotted orange line). Left and middle panels show data for uphill and downhill runs for strong gradient. The right panels show data for the zero gradient case. To clearly show the shift of the peak position with time, here we have presented the zoomed data near the peak region. Each data point has been averaged over at least $10^6$ histories. Other simulation parameters are as in Fig. \ref{fig:Dms}.} 
\label{fig:mtdist}
\end{figure}

\newpage
\section{Data for two dimensions}
\label{app:2d}
\renewcommand{\thefigure}{I\arabic{figure}} \setcounter{figure}{0}
In this section we present our simulation results for methylation dynamics for two dimensional motion of the cell. In a box of size $L_x \times L_y$, with reflecting boundary conditions at the four walls, a linear concentration profile $c(x)$ for the attractant is set up along the $x$-direction, and the $y$-direction has no gradient. Our measurement of $\Delta m^\pm (t)$ and $\delta m^\pm (t)$ show qualitatively similar behavior as seen in one dimension for the weak and strong gradient cases. Our data are presented in Figs. \ref{fig:Dmw2}, \ref{fig:dmw2}, \ref{fig:Dms2}, \ref{fig:dms2}. 
\begin{figure}
\includegraphics[scale=1.2]{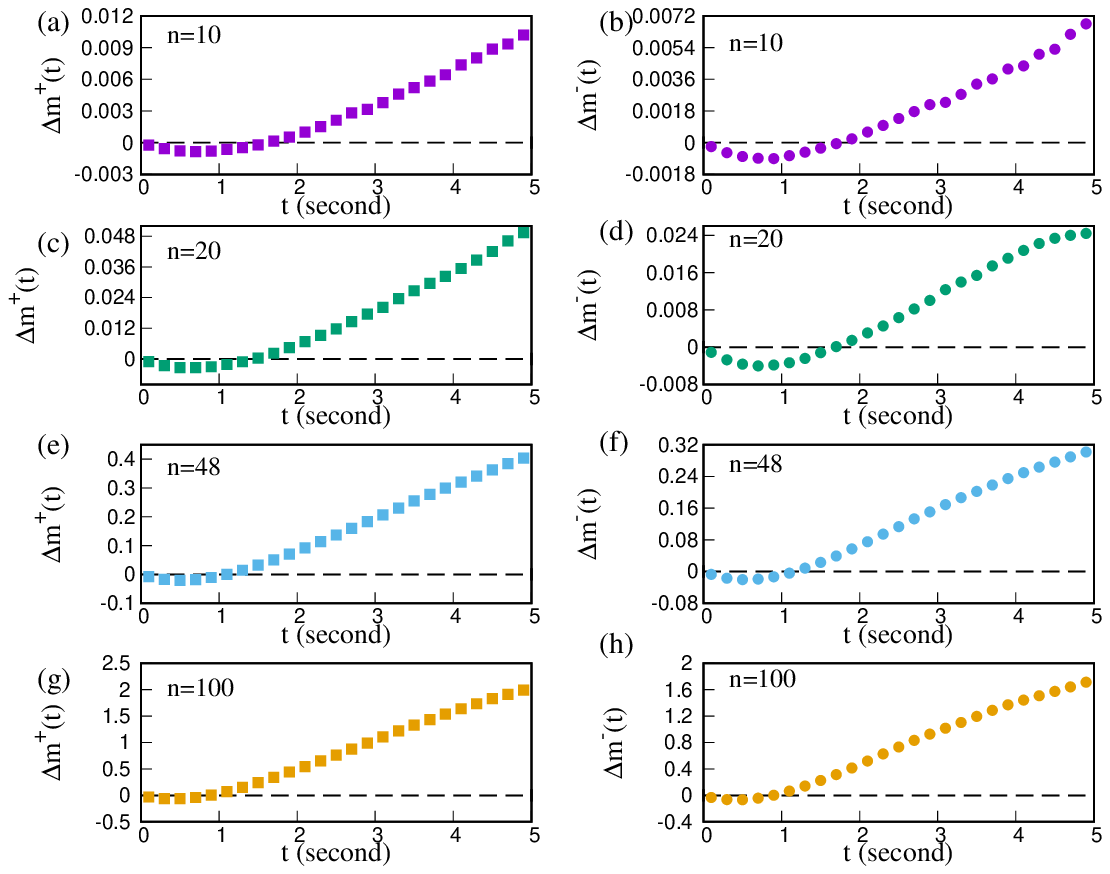}
\caption{Temporal variation of $\Delta m^\pm (t)$ for different $n$ in two dimensions: left panel shows plots for $\Delta m^+(t)$ for the uphill runs and the right panel shows $\Delta m^-(t)$ for the downhill runs. Here weak gradient of $c(x)$ is considered and the data look qualitatively similar to the one dimensional case, presented in Fig. \ref{fig:Dmw}. All simulation parameters are listed in Table \ref{table} in Appendix \ref{app:model}. These data are averaged over at least $10^5$ histories.}
\label{fig:Dmw2}
\end{figure}

\begin{figure}
\includegraphics[scale=1.2]{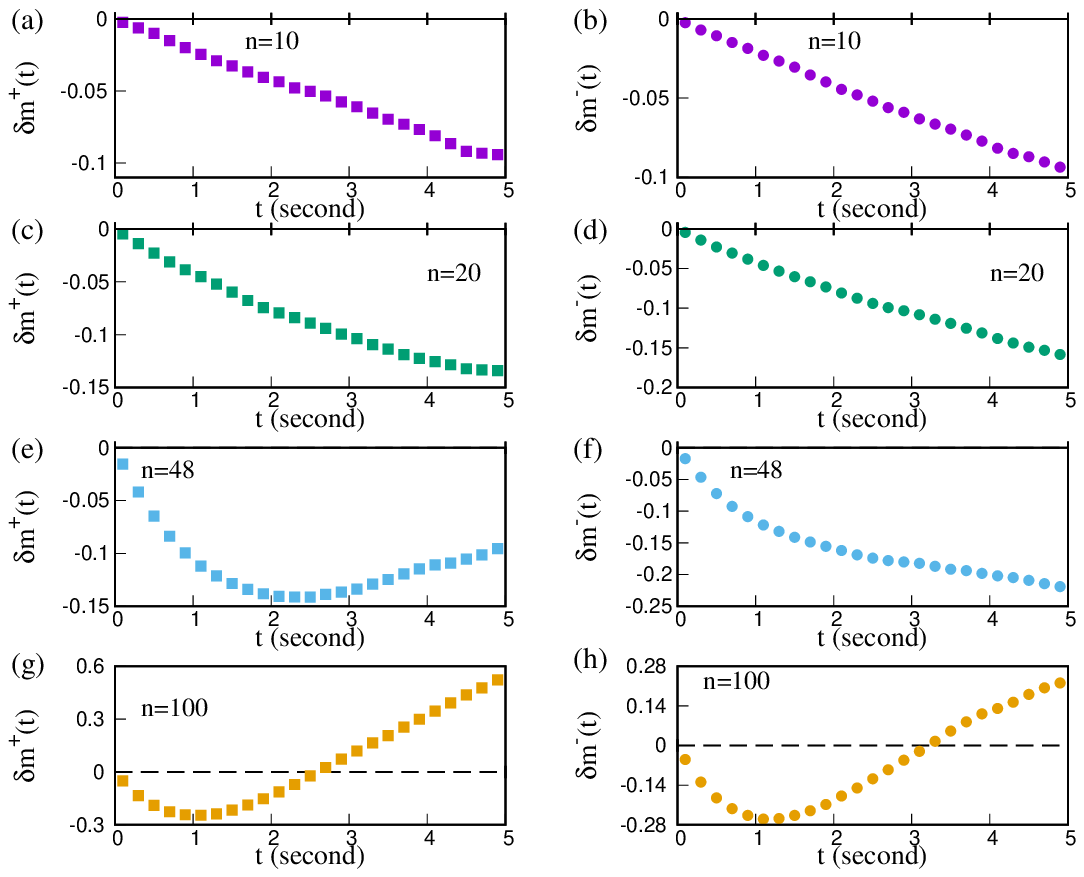}
\caption{Temporal variation of $\delta m^\pm (t)$ for different $n$ in two dimensions: left (right) column corresponds to uphill (downhill) runs. As seen in the data for one dimension, $\delta m^\pm (t)$ and $\Delta m^\pm (t)$ show opposite trends for small $n$, while for large $n$ their trends become similar. These data are averaged over at least $9 \times 10^5$ histories. Other simulation details are as in Fig.  \ref{fig:Dmw2}.}
\label{fig:dmw2}
\end{figure}

\begin{figure}
\includegraphics[scale=1.2]{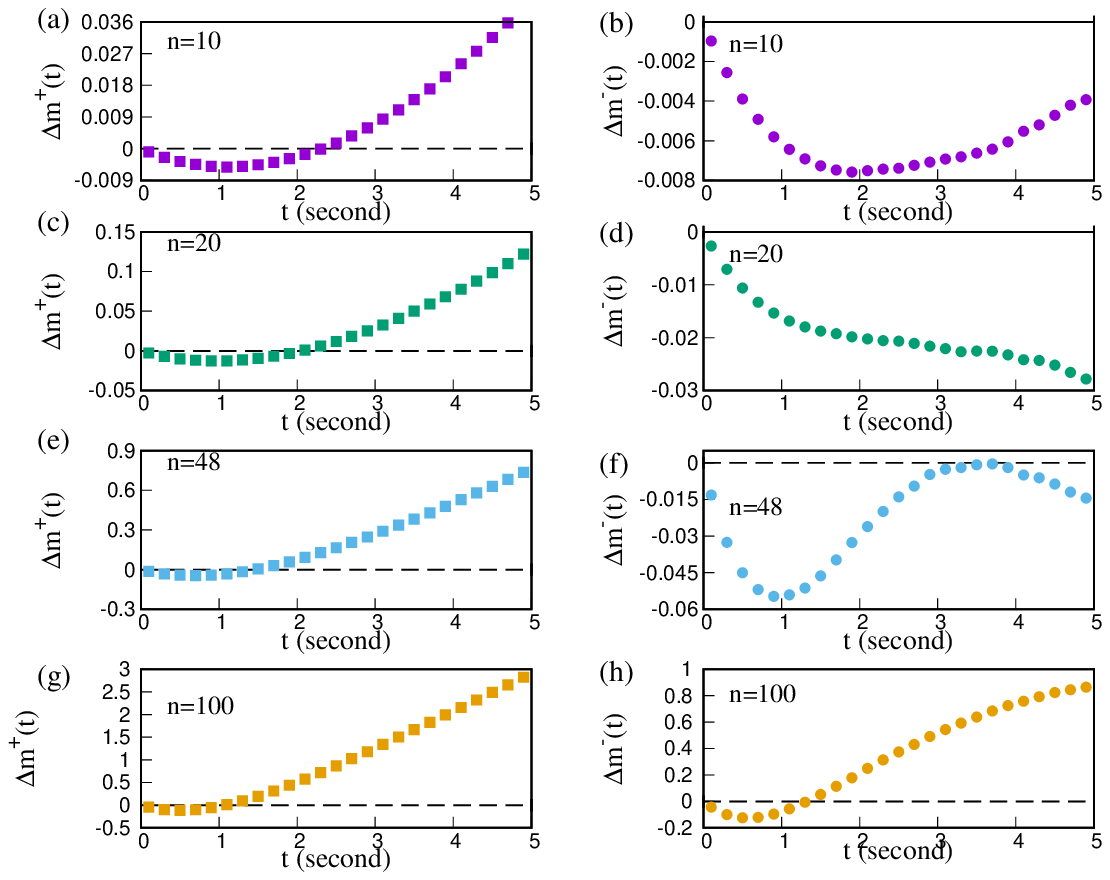}
\caption{Temporal variation of $\Delta m^\pm (t)$ for different $n$ and strong gradient case in two dimensions: left (right) column corresponds to uphill (downhill) runs. The qualitative nature of variation is similar to our data for one dimension in Fig. \ref{fig:Dms}. These data have been averaged over at least $10^5$ histories. } 
\label{fig:Dms2}
\end{figure}

\begin{figure}
\includegraphics[scale=1.2]{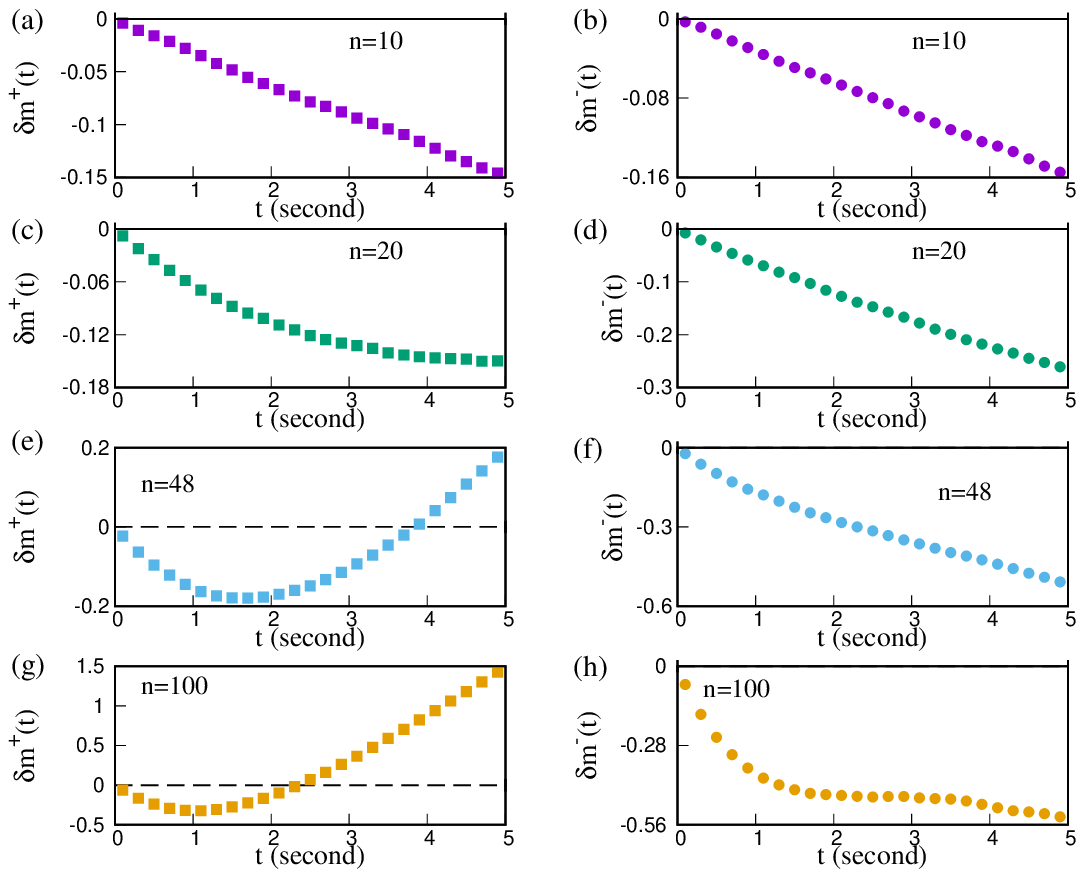}
\caption{Temporal variation of $\delta m^\pm (t)$ for different $n$ and strong gradient in two dimensions: left (right) column shows data for uphill (downhill) runs. The qualitative behavior is not too different from Fig. \ref{fig:dms} for the one dimensional case. These data are averaged over at least $7 \times 10^5$ histories.}
\label{fig:dms2}
\end{figure}

\end{document}